\documentclass{optica-article}

\journal{opticajournal} 

\articletype{Research Article}

\usepackage{lineno}
\usepackage{ulem} 

\begin{document}

\title{Explainable deep-learning detection of microplastic fibers via polarization-resolved holographic microscopy}

\author{Jan Appel,\authormark{1} Marika Valentino,\authormark{2} Lisa Miccio,\authormark{2,*} Vittorio Bianco,\authormark{2} Raffaella Mossotti,\authormark{3} Giulia Dalla Fontana,\authormark{3} Miroslav Ježek,\authormark{1} Pietro Ferraro,\authormark{2} Jaromír Běhal\authormark{1,*}}

\address{\authormark{1}Department of Optics, Palacký University, 17. listopadu 12, 77900, Olomouc, Czechia\\
\authormark{2}Istituto di Scienze Applicate e Sistemi Intelligenti ``Eduardo Caianiello'' (ISASI-CNR), via Campi Flegrei 34, 80078, Pozzuoli, Napoli, Italy\\
\authormark{3}STIIMA-CNR Institute of Intelligent Industrial Technologies and Systems for Advanced
Manufacturing, National Research Council of Italy, 13900 Biella, Italy;}

\email{\authormark{*}jaromir.behal@upol.cz, lisa.miccio@cnr.it} 


\begin{abstract*} 
Reliable identification of microplastic fibers is crucial for environmental monitoring but remains analytically challenging. We report {the first} explainable deep-learning framework for classifying microplastic and natural microfibers using {exclusively polarization-based features obtained from} polarization-resolved digital holographic microscopy. From multiplexed holograms, the complex Jones matrix of each fiber was reconstructed to extract polarization eigen-parameters describing optical anisotropy. Statistical descriptors of nine polarization characteristics formed a 72-dimensional feature vector for a total of 296 fibers spanning six material classes, including polyamide 6, polyethylene terephthalate, polyamide 6.6, polypropylene, cotton and wool. The designed fully-connected deep neural network achieved an accuracy of 96.7~\% on the validation data, surpassing that of common machine-learning classifiers. Explainable artificial intelligence analysis with Shapley additive explanations identified eigenvalue-ratio quantities as dominant predictors, revealing the physical basis for classification. An additional reduced-feature model with the preserved architecture exploiting only these most significant eigenvalue-based characteristics retained high accuracy (93.3~\%), thereby confirming their dominant role while still outperforming common machine-learning classifiers. These results establish polarization-based features as distinctive optical fingerprints and demonstrate the first explainable deep-learning approach for automated microplastic fiber identification.
\end{abstract*}

\section{Introduction}
Plastics have become pervasive in modern society, widely used in packaging, automotive
components, construction materials, electronics, and textiles. However, their resistance to
degradation and extensive post-consumer accumulation have raised serious environmental
concerns. Nowadays, microscopic polymer particles known as microplastics (in the size range
of 0.001–5 mm) are recognized as emerging pollutants \cite{Gesamp2020Sources}. These particles,
whether intentionally manufactured at microscopic scale (primary microplastics) or formed
through fragmentation of larger plastics (secondary microplastics), have been detected across
various environmental compartments, including marine and freshwater systems, soils, and even
the atmosphere. Their potential to enter food webs and accumulate in living organisms has prompted increasing attention to their potential environmental and health impacts \cite{Liu2025Recent, Yoganandham2023Micro}.

The global textile industry shift toward synthetic fibers contributes to a marked increase
in microfiber pollution. Indeed, a significant subset of microplastics originates from synthetic
micron-sized fiber fragments (MFFs) shed from textiles, accounting for 35~\% of the primary microplastics in the oceans
\cite{Mazhandu2020Integrated, Boucher2017Primary}. The physical characteristics of synthetic
MFFs, including small diameter, elongated shape, and low density, allow a substantial fraction to
bypass conventional wastewater treatment. As a result, they can be discharged into rivers, lakes,
and marine environments in significant quantities, which makes synthetic textile fibers a major
source of microplastic water contaminants \cite{Sun2019Microplastics, Henry2019Microfibres}.
These MFFs, typically composed of polyester, polyamide, or polypropylene, are released into the
environment during common processes such as laundering, wear, and mechanical abrasion.
Unlike natural fibers, synthetic microfibers are almost non-biodegradable and may retain
chemical additives, dyes, or finishing agents, enhancing their persistence and potential toxicity in
aquatic environments. Therefore, the identification and accurate classification of MFFs are
crucial for understanding why microplastics have become such pervasive pollutants. Such
insights are critical for advancing mitigation strategies and identifying weaknesses in the textile
production lifecycle, including washing, processing, recycling, and disposal, that contribute to
environmental contamination \cite{DeFalco2018Evaluation}.

Currently, no standardized method exists for microplastic identification and characterization. Instead, a range of analytical approaches are employed with their pros and cons, including visual identification, sieving, staining, gravimetric, spectroscopic, thermal analyses, and other methods \cite{Hildebrandt2019Evaluation, Lukose2025Gaining, Yan2024Pushing}. Visual methods are simple but operator-dependent and size-limited, although the specific staining with Nile red may categorize particles according to sample surface polarity \cite{Maes2017A}. Scanning electron microscopy coupled with energy dispersive X-ray spectroscopy is non-destructive and provide morphological and elemental analysis but the procedure is costly and requires cumbersome sample preparation, limiting number of samples that can be handled \cite{Shim2017Identification}. Methods based on analysis of the burning properties and solvent resistance, such as pyrolysis coupled with gas chromatography and mass spectrometry or thermal desorption gas chromatography–mass spectrometry, provide chemical signatures but are destructive thus making the sample unavailable for subsequent analysis \cite{Zarfl2019Promising}. Vibrational spectroscopies, such as attenuated total reflectance Fourier transform infrared (ATR-FTIR) \cite{Circelli2024Comparison}, focal plane array (FPA)-based micro FTIR \cite{Primpke2017An, Mintenig2017Identification}, Raman \cite{Seghers2021Preparation}, micro-Raman \cite{Cabernard2018Comparison}, and micro-FT-NIR \cite{Zhang2018Identification}, are widely used nondestructive polymer-specific identification techniques, though their require careful pretreatment because their performance may be degraded by interference with pigments, surroundings, organic matter, water, fluorescence signals, or low sample concentrations
\cite{Shim2017Identification, Zarfl2019Promising}.

Among label-free non-destructive optical methods, digital holography enables volume detection and characterization of microplastics in aqueous environments due to its numerical refocusing capability \cite{Nayak2021A, Takahashi2020Identification}. Quantitative phase imaging based on digital holography further provides the optical path difference, which is related to the morphology and refractive index of a sample. The technique was also combined with machine learning approaches \cite{Bianco2020Microplastic, Bianco2021Identification} to overcome the limited specificity of digital holography for classification enhancement. 
Alternative promising strategies exploit polarization-sensitive techniques such as polarized light microscopy \cite{Sierra2020Identification, Li2023Recognition} and polarization-resolved digital holography \cite{Jianqing2023Snapshot}.
{Indeed, lensless polarization-sensitive holographic systems have been introduced for automated, label-free microplastic analysis in flow \cite{Montandon2024}. Approaches such as computational polarized holography enable automated detection in scattering aquatic environments \cite{Huang2025}, while combined polarization and spectroscopic holography integrates multi-dimensional features for compositional analysis \cite{Zhu2024}. Polarization-resolved backscattering techniques exploit parameters such as the angle and degree of linear polarization, as well as Mueller matrix–derived features, to enable classification and differentiation of opaque and irregular microplastic particles \cite{Saur2026, Yang2026}.}
Finally, Jones phase microscopy \cite{Wang2008Jones, Jiao2020Real} enables material-specific eigen-polarization-based analysis of MFFs \cite{Behal2022Toward, Valentino2022Intelligent, Valentino2024Discern}. 

Distinguishing between different types of MFFs represents a typical multiclass classification task. Machine learning tools \cite{Owen2021Microplastic, DaSilva2020Classification, Kedzierski2019A} are nowadays applied to such problems, especially with high-dimensional data where intuitive rules for class assignment are rarely feasible \cite{Ballard2021Machine, Yuan2023Geometric, Freire2023Artificial, Bommasani2021On}. On the other hand, the nature of the data also guides the choice of learning paradigm, which in turn affects both methodology and interpretation of results. Especially, when samples are accompanied by explicit labels, supervised learning is typically employed, whereas in situations where only a portion of the data are labeled, semi-supervised learning can be utilized. On the contrary, when labels are absent altogether, unsupervised learning is applied; this does not result in explicit classification but rather in discovering natural groupings based on feature similarity. Moreover, it is also essential to distinguish between hard and soft classification, depending on the nature of the output decision. In hard classification, the model unambiguously assigns a sample to a single class. In contrast, soft classification provides probability distributions across classes, thus offering information about the degree of uncertainty in predictions made by the employed model.

Machine learning methods consequently differ in both their underlying principles and applicability to specific data types, respectively \cite{MLResearch_article, MLComparArticle}. The K-nearest neighbors is one of the most frequently applied algorithms, which classifies each sample according to the majority class among its K-nearest neighbors in the feature space. The naive Bayes classifier employs Bayes theorem under the assumption of feature independence, estimating the class to which a sample most likely belongs. Decision trees operate by sequentially partitioning data according to rules based on individual attribute values, while their more robust extension, random forests, aggregate the outcomes of multiple trees for improved performance with sufficient robustness against overfitting. The support vector machine method seeks to find the optimal separating hyperplane with the maximal margin between classes, and it is effective for classification with limited datasets. Logistic regression is often used for binary classification, modeling class membership probability via the logistic function. Boosting methods such as gradient boosting iteratively build an ensemble of weak classifiers, each correcting the errors of its predecessors, making them effective for use with limited datasets. Finally, artificial neural networks, especially when implemented in deep learning architectures, utilize layers of interconnected neurons to model complex, non-linear relationships \cite{Cybenko1989, Lu2017}. Each of these techniques exhibits specific strengths and limitations regarding interpretability, classification accuracy, robustness to noise, and computational requirements.

In particular, deep neural networks have been implemented for enhanced microplastic classification {using multimodal features that combine morphological and chemical data \cite{Takahashi2023}, for shape-based classification from scanning electron microscopy images \cite{Shi2022}, and for classification from holograms via a zero-shot learning approach \cite{Zhu2022}. They have also been used to distinguish microplastics from organic materials and to classify microplastics using Raman spectra \cite{Lee2023Automatic, Zhang2023A}.}
Other convolutional neural network models \cite{Gkioxari2017Mask, Akkajit2023Comparative} have been applied to classify images of large marine microplastics \cite{Han2023Deep}. {Recently, specialized architectures like vision transformers exploiting polarization images were also implemented \cite{Guo2026}.} Finally, deep learning has been combined with digital holography, to overcome its limited specificity, yielding improved classification performance by exploiting raw interference patterns \cite{Zhu2021Microplastic} as well as reconstructed amplitude and phase masks through image fusion techniques \cite{Russo2024Deep}.

In this work, we present a non-destructive label-free {material-specific MFF classification method based on a deep neural network that uses only polarization-derived features, achieving} unprecedented accuracy. In particular, we analyze four types of common synthetic MFFs, including polyamide 6 (PA 6), polyethylene terephthalate (PET), polyamide 6.6 (PA6.6), polypropylene (PP), and two types of common natural MFFs (cotton and wool) using polarization-resolved digital holographic microscope. The retrieved eigen-parameters calculated from the measured polarization Jones matrix of the sample are related to the MFF geometry. {The resulting set of quantities forms the feature space used in the subsequent classification pipeline}. We show that these {polarization-based} features enable efficient classification with the {fully-connected} deep neural network, achieving an accuracy of 96.7\% accuracy {and demonstrating that deep learning with a large, overparameterized network can be successfully applied to small real-world microfiber datasets, outperforming the current state of the art. Furthermore, we provide the first explainability analysis of a deep learning model in this application by employing the} SHAP framework to identify the polarization and statistical features {that most strongly contribute to classification performance}. 

\section{Measurements and feature extraction}
This section describes setup used for polarization-sensitive holographic imaging as well as Jones matrix reconstruction, and extraction and meaning of MFF polarization characteristics determining features considered for subsequent classification pipeline.

{Common imaging geometries suitable for digital holographic microscopy with diffraction-limited imaging quality are a telecentric configuration (microscope objective with a tube lens) and a finite-distance objective configuration \cite{Sirico2022l}. In both cases, proper compensation of phase-background distortions, especially phase curvature, is essential to maintain separation of holographic orders and preserve resolution \cite{Sanchez-Ortiga2014}. The finite-distance approach typically exhibits residual tilt and curvature (with curvature pre-compensated experimentally), which simplifies phase unwrapping and phase-background fitting to a one-dimensional problem \cite{Zuo2013}. The telecentric configuration reduces phase curvature under well-collimated conditions and simplifies phase-curvature correction, but imposes constraints on additional optics and alignment precision. In practice, alignment tolerances increase the complexity of the resulting phase-background structure, which may require full two-dimensional phase-background fitting approach, if feasible. Therefore, we adopted the finite-distance approach for its simplicity and flexibility. This geometry also allows a shorter sample–camera distance and enables reference-beam mixing using smaller beam splitters without aperture truncation. Nevertheless, numerical phase-background correction remains an integral part of the processing pipeline in both approaches, as it is necessary for accurate complex-amplitude refocusing and quantitative phase analysis.}

\begin{figure}
  \centering
  \includegraphics[width=1\textwidth]{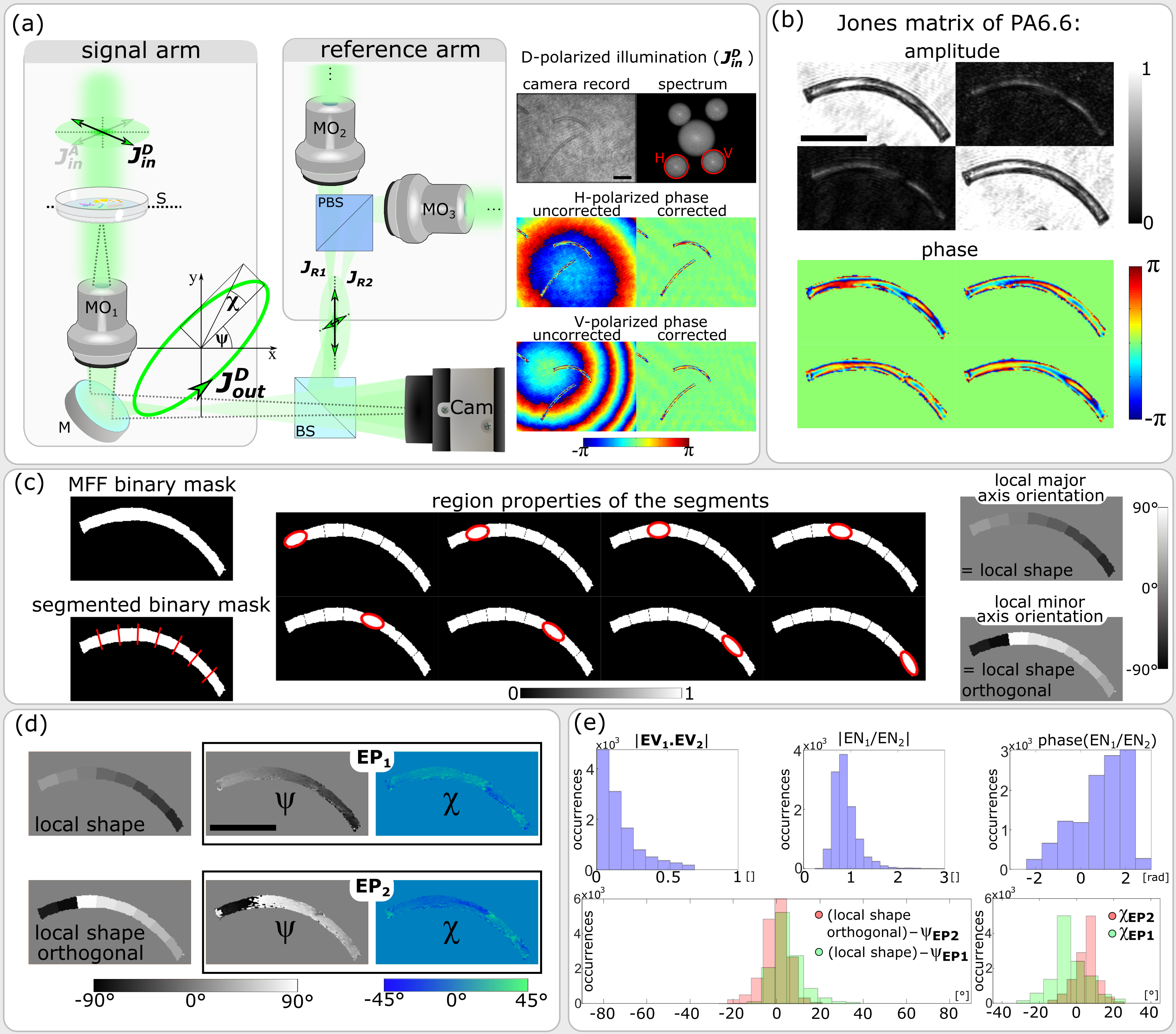}
  \caption{(a) Simplified sketch of experimental configuration {with an example of an interference pattern, multiplexed Fourier spectrum, and uncorrected and corrected phase maps, which correspond to horizontally and vertically polarized components for the diagonally polarized sample illumination $J_{in}^D$}. (b) Example of retrieved Jones matrix for PA6.6 MFF. {(c) Local shape and local shape orthogonal representing major and minor axis orientation of the segmented binary mask.} (d) Local shape map and map orthogonal to the local shape compared to parameters of eigenpolarizations $EP_1$ and $EP_2$, respectively. (e) Histograms of the derived polarization characteristic. {In particular, the total number of 11,968 evaluated pixels provides sufficient data for accurate statistical analysis of the PA6.6 MFF.} Black scale bars in (a), (b) and (d) represent 100 $\mu$m lengths.}
  \label{fgr:MeasureSketch}
\end{figure}

Simplified sketch of the realized digital holographic microscope based on Mach-Zehnder interferometer is shown in Fig.~\ref{fgr:MeasureSketch}(a). The initial spatially filtered, collimated, and polarized laser beam (Sapphire SF, wavelength 532 nm) is split and directed into three independent optical paths, i.e., a signal arm, a horizontally polarized reference arm, and a vertically polarized reference arm, respectively. The studied sample present in the signal arm is illuminated by diagonally (or antidiagonally) linearly polarized beam. The sample is imaged by a microscope objective MO$_1$ (5$\times$/0.12) onto a camera (UI 1550SEC-HQ, 2.8 $\mu$m square pixels) with lateral magnification 6.6. The mutually slightly tilted horizontally and vertically polarized reference beams $J_{R1}$ and $J_{R2}$ are combined via a polarizing beam splitter PBS and further injected into the signal arm via a beam splitter BS. All three mutually coherent beams interfere and create a multiplexed interference pattern recorded by the camera; hence a single-shot polarization state reconstruction is feasible. The microscope objectives MO$_2$ and MO$_3$ inserted in both reference arms reduce phase curvatures in the reconstructed phase profiles. 

{Design} of the holographic microscope allows reconstruction of complex amplitudes corresponding to horizontal and vertical polarization-state components, which are obtained via spectral filtering of a multiplexed interference pattern. 
{An example of the resulting interference pattern for diagonally polarized sample illumination $J_{in}^D$ is shown in Fig.~\ref{fgr:MeasureSketch}(a), together with the multiplexed Fourier spectrum, from which the complex amplitudes of the horizontally and vertically polarized components are reconstructed. First, the spatial carrier of a selected spectral order is removed, and all other spectral components are suppressed. The corresponding complex amplitude is then obtained via a two-dimensional inverse Fourier transform. This procedure is repeated for the second diffraction order, yielding the complex amplitudes of both polarization components. It can be noticed that the associated quantitative phase maps exhibit phase-background distortions, primarily due to residual tilt and curvature. These contributions are removed using a reference interference record acquired in a sample-free region prior to measurement \cite{Behal2019}, followed by one-dimensional phase-background fitting to correct residual errors \cite{Zuo2013}. The amplitudes are normalized to the same reference. Finally, for a known input polarization $J_{in}^D$, the polarization-state distribution $J_{out}^D$ at the detector plane is obtained from the reconstructed complex amplitudes. An analogous procedure is applied for $J_{in}^A$ to obtain $J_{out}^A$.} Assuming that $J_{in}^D$ and $J_{in}^A$ are orthogonally polarized, the $2\times2$ complex-amplitude Jones matrix describing the sample polarization response can be calculated directly \cite{Behal2022Toward}, {see Fig.~\ref{fgr:MeasureSketch}(b)}.

Once the Jones matrix of the sample is retrieved its properties based on eigen-analysis can be evaluated. In Jones-matrix formalism it is often convenient to extract eigenvectors (EVs) and their associated eigennumbers (ENs) to better understand polarization properties of the sample. The EVs correspond to polarization modes that propagate throughout samples unaltered and are therefore described by the orientation angle $\psi$ and the ellipticity angle $\chi$; hence, the EVs are also referred to as eigenpolarizations (EPs). Each of the EVs is modulated in both amplitude and phase by its corresponding EN. Thus, the magnitude of the ENs’ ratio indicates the degree of anisotropic absorption, while the phase of this ratio reveals the relative phase delay between the two EVs \cite{Baroni2020Extending}. Furthermore, the inner product of the EVs provides a measure of polarization homogeneity of the sample \cite{Lu1994Homogeneous}.

Initially, for each MFF a binary mask was calculated, which served as a reference for orientation-dependent polarization properties. For this purpose, the binary mask was further divided into disjunct {segments. The region properties of each segment were calculated, including the orientation of the major axis, which locally approximates the tangent to the MFF. The conjunction of all the segments with their corresponding major-axis orientation angles consequently defines the local orientation along the MFF tangent. We refer to this approximation as the local shape, consult with Fig.~\ref{fgr:MeasureSketch}(c). Similarly, the minor axis of each segment locally approximates the normal to the MFF. The conjunction of all segments with their corresponding minor-axis orientation angles defines the local orientation of the MFF normal. We refer to this approximation as the local shape orthogonal, as its orientations are orthogonal to those of the local shape defined above, see Fig.~\ref{fgr:MeasureSketch}(c).}
Indeed, a clear similarity between the local shape direction and the orientation angle $\psi$ of $EP_1$ ($\psi_{EP1}$) in Fig.~\ref{fgr:MeasureSketch}(d) justifies the use of the local shape map as a reference for $\psi_{EP1}$. Similarly, the direction orthogonal to the local shape serves as a reference for $\psi_{EP2}$, i.e., the orientation angle of $EP_2$. Calculations of EP orientations thus show that $EP_1$ aligns with the MFF’s principal axis, while $EP_2$ tends to be approximately perpendicular Fig.~\ref{fgr:MeasureSketch}(d).

For each analyzed MFF we calculated several polarization-based quantities. The considered quantities include the magnitude of the EVs inner product $(\left| EV_1\cdot EV_2\right|)$, and the modulus, phase, real part, and imaginary part of the ENs ratio (i.e., $(\left| EN_{1}/EN_{2}\right|)$, phase($EN_{1}/EN_{2}$), Re($EN_{1}/EN_{2}$), Im($EN_{1}/EN_{2}$)), as well as the ellipticity angles $\chi$ of $EP_1$ and $EP_2$ (i.e., $\chi_{EP1}$ and $\chi_{EP2}$). Moreover, the polarization ellipse main axis orientation angles $\psi$ of $EP_1$ and $EP_2$ (i.e., $\psi_{EP1}$ and $\psi_{EP2}$) were also calculated and related to the MFF geometry via the local shape and the orientation perpendicular to the local shape, respectively. All these nine quantities will be referred to as polarization characteristics throughout the following text.

{Each polarization characteristic, evaluated in a pixel-wise manner, is sampled over a large number of pixels within the MFF image and therefore exhibits a statistical distribution. Representative histograms for PA6.6 MFF are shown in Fig.~\ref{fgr:MeasureSketch}(e). Accordingly, we calculated several statistical parameters of these polarization characteristics, including the mean, median, mode, mean absolute deviation, median absolute deviation, standard deviation, skewness, and kurtosis. The evaluation was restricted only to regions with sufficient amplitude to exclude low-amplitude pixels while retaining enough data for statistical analysis. Low-amplitude pixels can introduce irrelevant contributions and artificially broaden the distributions of polarization characteristics, as their information is more strongly affected by noise. In addition, the selection criterion must therefore ensure a sufficient number of pixels for reliable statistical evaluation, simultaneously. In our case, a threshold of 35~\% of the average background amplitude was simply applied, to be in accordance with the reference \cite{Behal2022Toward}, resulting in thousands of evaluated pixels per sample. The median number of evaluated pixels was 5,447 for PA6, 15,211 for PET, 8,589 for PP, 9,569 for PA6.6, 11,387 for cotton, and 13,070 for wool.}

To sum up, we calculated 8 statistical parameters for 9 polarization characteristics, creating a unique set of 72 numbers for each MFF, which we call features throughout the following text. In total, we analyzed 296 MFFs, prepared according to previously developed protocols \cite{Behal2022Toward}, including 47 MFFs of PA6, 51 of PET, 46 of PP, 47 of PA6.6, 47 of cotton, and 58 of wool. Therefore, the input data used for the subsequent classification comprised of 296 measured MFFs belonging to six distinct classes and each analyzed MFF consisted of 72 features and a discrete label representing its class.  

\section{Neural network architecture}

The following section details each step of the classification pipeline, including data preparation and normalization, data partitioning into training and validation sets, neural network design, and model accuracy evaluation.

The process of neural network training typically requires the division of the available dataset into at least two subsets: a training set and a validation set. In this study, the dataset of MFFs was chosen such that 80~\% of the MFFs were allocated to the training set and 20~\% to the validation set. This approach ensures development of a robust model with reliable generalization performance. Moreover, normalization of both input and output data is crucial for stable and efficient neural network learning. It transforms values to a common scale, such as mapping to [0,1] or applying zero mean and unit variance (Z-transformation). Importantly, the same procedure must also be applied to validation data to ensure comparability of results and proper model performance on unseen data. In our case, we used Z-transformation. To address the classification task a fully-connected (dense) neural network architecture was selected, which is a well-suited approach for modeling complex, non-linear relationships in such data. The use of fully-connected network is justified by the tabular nature of the input data, where each MFF is represented by a fixed-length feature vector without inherent spatial or sequential structure.

\begin{figure}
  \centering
  \includegraphics[width=0.70\textwidth]{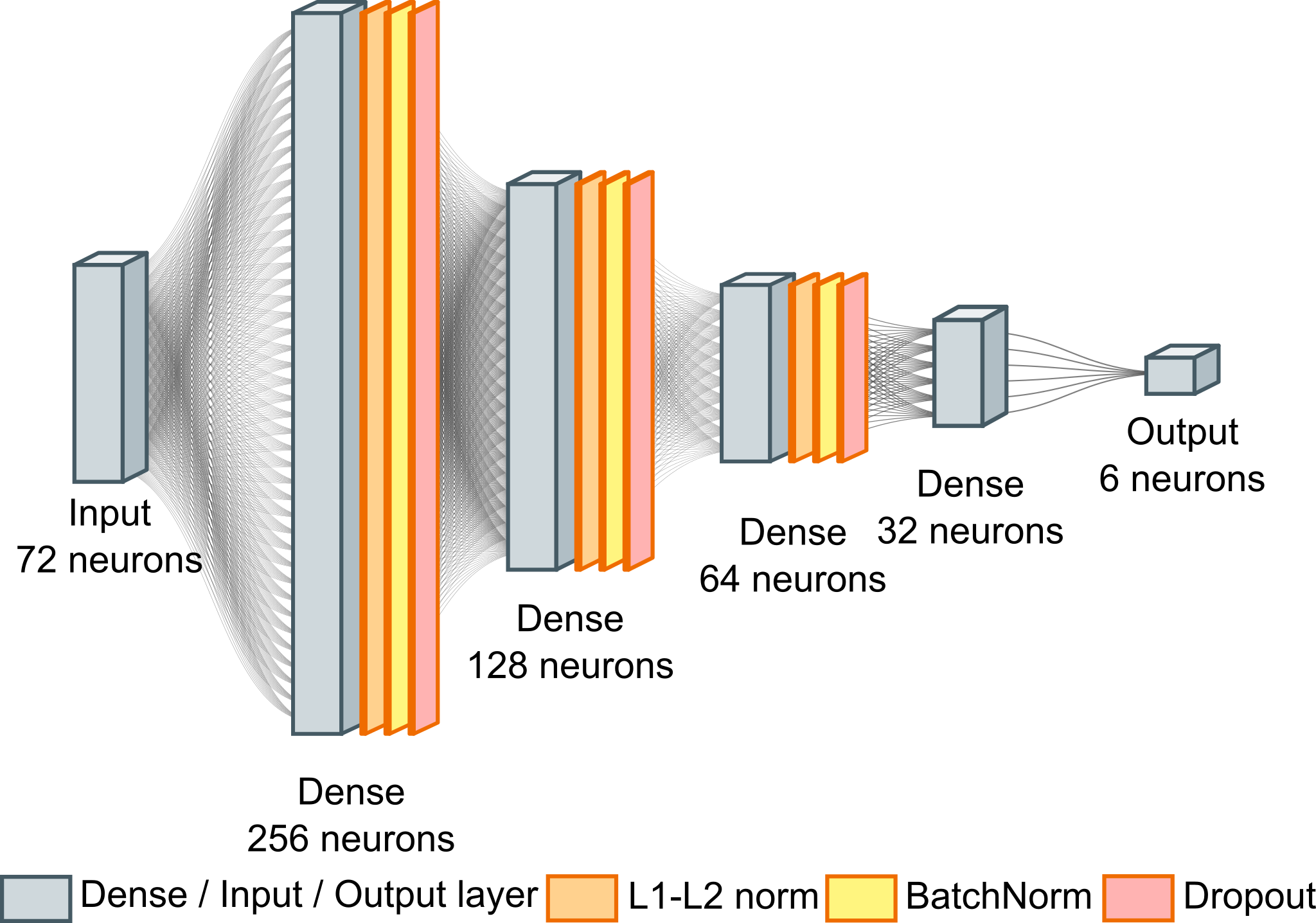}
  \caption{{Architecture of the implemented deep} neural network. Input layer has dimensions corresponding to the number of features. Four hidden layers with a progressively decreasing number of neurons. All hidden layers in the network employ regularization techniques, specifically: batch normalization, dropout and additional regularization methods, see text for details.}
  \label{fgr:DLmodel}
\end{figure}

The training process was conducted in a supervised learning regime, utilizing labeled data to optimize the network parameters. To account for class imbalance in the dataset, class weights were incorporated during training. The implemented {fully-connected} neural network can be divided into three parts, see Fig. ~\ref{fgr:DLmodel}. {The first layer inputs the feature} vector with 72 elements. Second part is composed of four hidden layers, each designed to progressively extract higher-level representations. Layers contain a progressively decreasing number of neurons: 256, 128, 64, and 32, respectively. This funnel-shaped architecture provides efficient abstract feature extraction. At the same time, the gradual reduction in the number of neurons leads to less computational effort. Finally, the output layer provides the final probabilities for each of the six MFF classes. Total number of trainable parameters is 63 thousands.

The first layer and all hidden layers have a rectified linear unit (ReLu) activation function. The last layer has a Softmax activation function due to its probabilistic interpretation suitable for multi-class classification \cite{Goodfellow2016Deep}. To reduce overfitting and enhance generalization, architecture employed regularization techniques such as batch normalization and dropout layers \cite{Srivastava2014Dropout} with relatively high dropout rates in first hidden layer (0.8) and gradually decreasing through the second hidden layer (0.4) to the third hidden layer (0.2). These layers also contain $L_1$ and $L_2$ regularizations. Their specific use ($L_1$, $L_2$, or both) and values vary depending on the layer and its part (kernel, bias, and activity regularizer). The last hidden layer does not contain any regularization technique. Training was performed using the Adam optimizer \cite{Kingma2015A} (initial learning rate 0.001) with categorical cross-entropy loss, and performance was monitored via categorical accuracy. The model was trained for up to {300} epochs with a batch size of 50. A detailed layer-by-layer description of the architecture, including all regularization parameters, training callbacks, and the class weighting strategy, is provided in repository \cite{repository}.

\section{Classification performance}
{The suitability of the model architecture was assessed by monitoring the training and validation learning curves. The validation accuracy gradually converges to the high training accuracy, indicating good generalization, as the model performs consistently on both datasets. No evidence of overfitting is observed, since the validation accuracy does not degrade while the training accuracy approaches its maximum. These results suggest that the model captures meaningful patterns while maintaining robust generalization. In particular, it may appear counterintuitive that a relatively large model (63,000 trainable parameters) does not overfit when trained on a comparatively small dataset (236 training samples). As discussed above, multiple regularization techniques were employed to stabilize training. Furthermore, on the general note, there is well documented double descent behavior of deep neural networks \cite{Belkin2019, Nakkiran2021}, where the overfitting might be present for smaller models but disappears for heavily overparametrized models, i.e., when the complexity of the network is significantly larger than the size of training dataset.}

\begin{figure}
  \centering
  \includegraphics[width=1\textwidth]{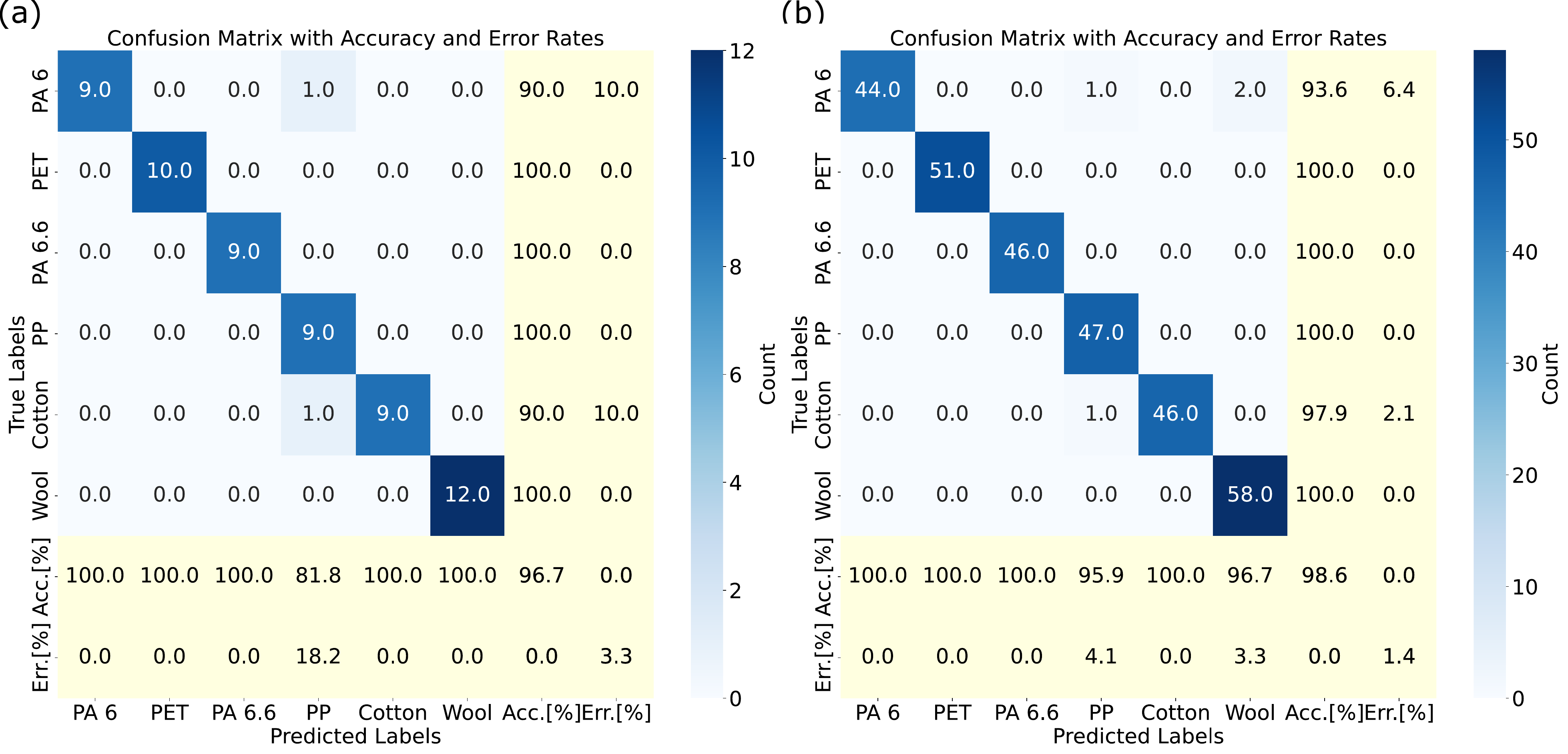}
  \hspace{15mm}
  \caption{Confusion matrix of the model trained on labeled data. (a) Evaluation on the validation data (60 labeled MFFs). (b) Evaluation on the complete dataset of training and validation data (296 labeled MFFs).}
    \label{fgr:ConfMatrix}
\end{figure}

To assess the performance of the trained model in detail, a confusion matrix approach is employed. For comparative evaluation, the confusion matrix was computed on the entire dataset, including the training samples, as it is sometimes considered to evaluate the classification performance with low number of samples, see  Fig.~\ref{fgr:ConfMatrix}(b). However, since assessing model accuracy on a dataset that includes training data may provide an overly optimistic estimate of generalization performance, we also provide the confusion matrix for the validation data only, see Fig.~\ref{fgr:ConfMatrix}(a). MFF classes in Fig.~\ref{fgr:ConfMatrix} are labeled as PA 6, PET, PA6.6, PP, cotton, and wool.

The confusion matrix in Fig.~\ref{fgr:ConfMatrix}(a) summarizes the classification performance of the current model on the validation dataset consisting of 60 labeled MFFs. It provides a detailed breakdown of true versus predicted labels, including class-wise accuracy and error rates. Specifically, only two misclassifications appear among the MFF classes, both being false negatives for PP. In these cases, the model misclassified MFFs from PA 6 and cotton as PP. The diagonal values correspond to the number of true positives, which allows quantification of model accuracy. This accuracy is calculated by dividing their sum by the total number of predictions, yielding 96.7~\% for our model, an unprecedented value in the field.

The confusion matrix in Fig.~\ref{fgr:ConfMatrix}(b) summarizes the classification performance of the current model on the dataset containing all 296 labeled MFFs, i.e., the complete dataset including MFFs used for training. Across these 296 predictions, only four misclassifications occurred. Since the dataset contains complete data from the previous evaluation, two of these errors origin from the previous validation dataset—false positives for PA 6 and cotton, both misclassified as PP. In addition, two false negatives arise from the training data for PA 6: one misclassified as PP and the other as wool. The overall model accuracy on this dataset reaches 98.6~\%.

{
To assess the possible presence of feature redundancy we calculated Pearson correlation matrix for all 72 features. Based on this analysis, we identified majority of features with correlation coefficient value between -0.8 and 0.8, indicating that most of the features are not redundant. Hence, strict feature selection based on value of correlation coefficient could negatively affect the classification accuracy, as important features could be lost. Especially, the upper triangular matrix without considering diagonal of the Pearson matrix contains 2556 elements from which only 145 have correlation coefficient outside of this range, i.e. 5.7\%.
Nevertheless, it is important to note that feature correlation and classification performance describe different properties of a dataset. While reducing feature correlation may help mitigate redundancy, it does not necessarily improve classification power. Ultimately, classification performance depends on the ability of a model to learn true, heavily nonlinear decision boundaries in the underlying (unknown) feature space, for which deep neural networks seem to be the optimal solution due to their universal approximation ability \cite{Cybenko1989, Hornik1989}. In other words, classification performance must be quantified.}

\begin{table}
  \centering
    \caption{Comparison of different classifiers. Values are reported as without feature selection / with feature selection.}
    \label{tbl:class_compare}
    \begin{tabular}{lcccc}
    \hline
    \rowcolor{gray!20}
    \textbf{Classifier} & \textbf{Accuracy} & \textbf{Precision} & \textbf{Recall} & \textbf{F1 score} \\
    \hline
    Gaussian naive Bayes   & 0.80 / 0.82 & 0.81 / 0.83 & 0.80 / 0.82 & 0.80 / 0.82 \\
    random forest          & 0.83 / 0.85 & 0.84 / 0.86 & 0.83 / 0.85 & 0.83 / 0.85 \\
    K-nearest neighbors    & 0.87 / 0.90 & 0.88 / 0.91 & 0.86 / 0.90 & 0.86 / 0.89 \\
    support vector machine & 0.88 / 0.90 & 0.88 / 0.91 & 0.88 / 0.90 & 0.88 / 0.90 \\
    gradient boosting      & 0.88 / 0.88 & 0.87 / 0.87 & 0.87 / 0.87 & 0.87 / 0.87 \\
    logistic regression    & 0.90 / 0.92 & 0.90 / 0.92 & 0.90 / 0.92 & 0.90 / 0.92 \\
    {fully-connected} &  &  & & \\[-0.7em]
    deep neural network    & 0.97  & 0.97  & 0.97  & 0.97 \\[-0.7em]
    (the main result of our work) &  &  & & \\
    \hline
    \end{tabular}
\end{table}

{Precisely for these reasons, several classifiers without and with feature selection were further tested.} Especially, machine learning algorithms, including Gaussian naive Bayes, random forest, support vector machine, K-nearest neighbors, gradient boosting, and logistic regression, were systematically implemented in addition to the primary classification framework. Moreover, beyond algorithm selection, it is useful to assess multiple metrics for quick yet effective evaluation of classifier performance. While the confusion matrix enables detailed error analysis, metrics such as accuracy (returns the fraction of correctly classified samples), precision (ratio $t_p / (t_p + f_p)$, where $t_p$ is the number of true positives and $f_p$ the number of false positives), recall (ratio $t_p / (t_p + f_n)$, where $f_n$ is the number of false negatives), and F1 score (2$\cdot$precision$\cdot$recall/(precision+recall), which represents harmonic mean of precision and recall) offer a simplified assessment. Hence, for each technique, the attained metrics were evaluated specifically on the validation dataset, ensuring that the reported results objectively reflect generalization performance and are not biased by the training process. 
The achieved results are summarized in Table~\ref{tbl:class_compare}, showing that the performance of these classifiers is inferior to the developed {fully-connected} deep neural network, even when they are combined with a {minimum redundancy maximum relevance (mRMR) feature selection algorithm} based on mutual information \cite{Peng2005, Radovic2017}. {More details provided in repository \cite{repository}.}

\section{Explainability}
We employ Shapley additive explanations (SHAP) to analyze the contribution of polarization-derived features used in the classification of MFFs. SHAP provides a principled, game-theoretic framework that decomposes each model prediction into additive contributions from individual input features relative to a baseline prediction\cite{Lundberg2017, Lundberg2020, PonceBobadilla2024}. To obtain a global measure of relevance, we report SHAP feature importance (SHAP-FI) values, defined as the average absolute SHAP values across all validation data points. A SHAP-FI value quantifies how strongly a particular feature, on average, shifts the model output in its natural prediction scale away from the baseline. This allows us to directly compare the relative influence of polarization characteristics and their statistical parameters on classification performance, thereby identifying which aspects of the polarization response carry the greatest predictive power for distinguishing between different MFFs.

\begin{figure}
  \centering
  \includegraphics[width=1\textwidth]{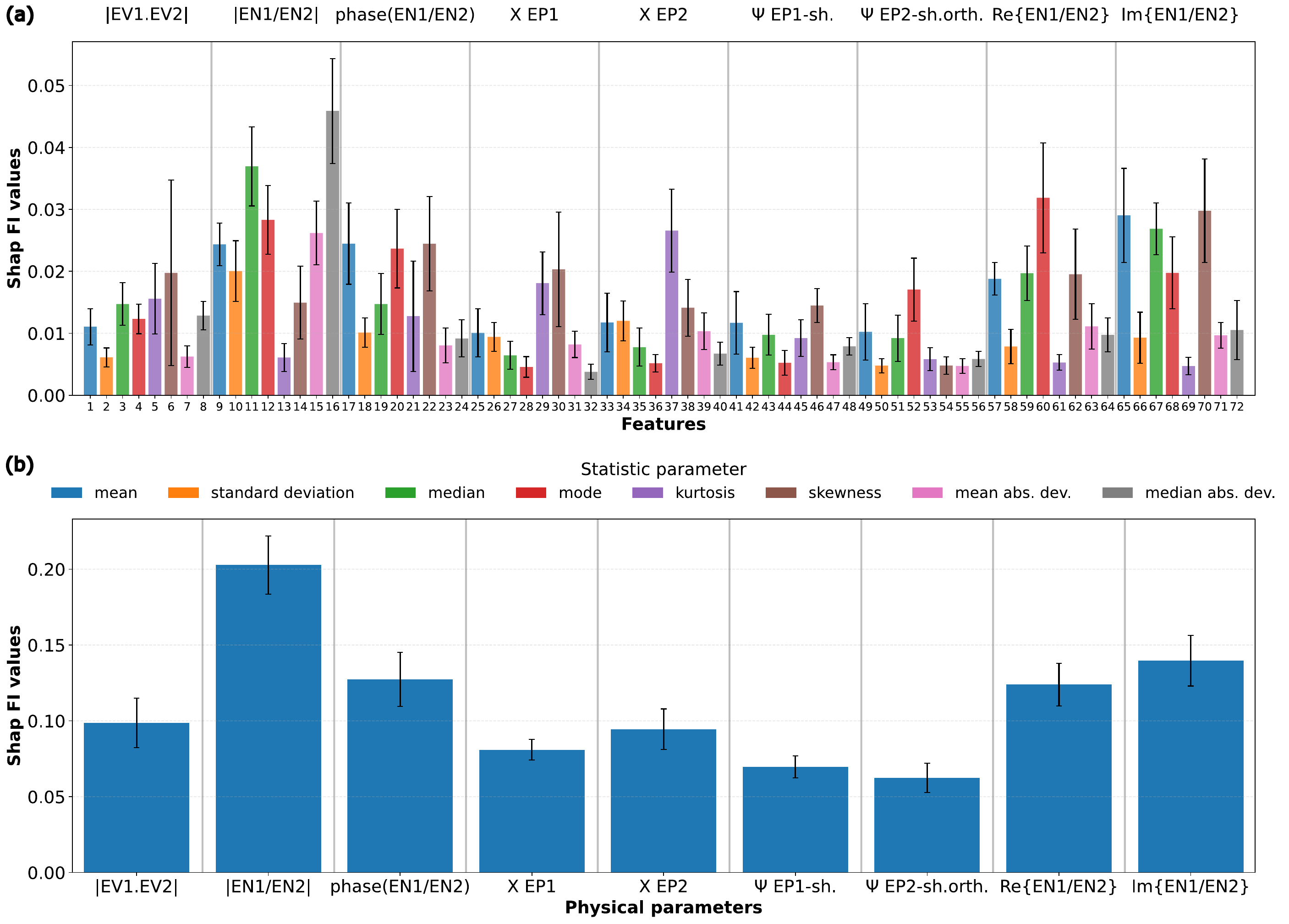}
  \caption{{SHAP-FI analysis.} (a) SHAP-FI values of nine polarization characteristics and their eight corresponding statistical parameters, yielding 72 SHAP-FI values. (b) Sums of the SHAP-FI values for the current polarization characteristics across all eight statistical parameters. $\left|EV_1\cdot EV_2\right|$ - magnitude of the EVs inner product; $\left|EN_{1}/EN_{2}\right|$, phase$(EN_{1}/EN_{2})$, Re$(EN_{1}/EN_{2})$, Im$(EN_{1}/EN_{2})$ - modulus, phase, real part, and imaginary part of the ENs ratio; $\chi_{EP1}$, $\chi_{EP2}$ - ellipticity angles of $EP_1$ and $EP_2$; $\psi_{EP1}$-sh. - orientation of $EP_1$ relative to the local shape; $\psi_{EP2}$-sh.orth. - orientation of $EP_2$ relative to the direction perpendicular to the local shape.}
  \label{fgr:Shap}
\end{figure}

Nevertheless, SHAP-FI values are not unique for repeatedly retrained models with the same parameters if the features are even slightly correlated or redundant \cite{Lundberg2020,Frye2020,Molnar2022}. For this reason, we performed an analysis of SHAP-FI values for twenty-eight trained models with the same architecture, accuracy, and confusion matrix. The calculated SHAP-FI values of each model were normalized with respect to their sum over the validation data. Mean values and standard deviations of the SHAP-FI values are shown in Fig.~\ref{fgr:Shap}(a). The most significant polarization characteristic is the absolute value of the ENs ratio, i.e. $\left|EN_{1}/EN_{2}\right|$, with the sum of mean SHAP-FI values across all eight statistical parameters equal to 0.203, which can be compared to the overall sum of all SHAP-FI values of 1. Among them, the median absolute deviation (0.046) and the median (0.037) are simultaneously the highest of all 72 SHAP-FI values. In addition, Fig.~\ref{fgr:Shap}(b) shows the sums of SHAP-FI values for the current polarization characteristics across all eight statistical parameters; ordered by to their significance the sums reach as follows: 0.203 for $\left|EN_{1}/EN_{2}\right|$, 0.140 for Im$(EN_{1}/EN_{2})$, 0.127 for phase$(EN_{1}/EN_{2})$, 0.124 for Re$(EN_{1}/EN_{2})$, 0.099 for $\left|EV_1\cdot EV_2\right|$, 0.094 for $\chi_{EP2}$, 0.081 for $\chi_{EP1}$, 0.070 for $\psi_{EP1}$-sh., and 0.062 for $\psi_{EP2}$-sh.orth. {The small variability of SHAP-FI values shows that the feature correlation and redundancy are relatively small in our case.} It is worth noting that the first four most significant polarization characteristics are based on ENs only (sum 0.594) and not on EVs and EPs (sum 0.406). To be complete, the difference between the SHAP-FI values for statistical parameters (summed across polarization characteristics) is less pronounced than between the polarization characteristics (summed across statistical parameters), which were already discussed.

\begin{figure}
  \centering
  \includegraphics[width=1\textwidth]{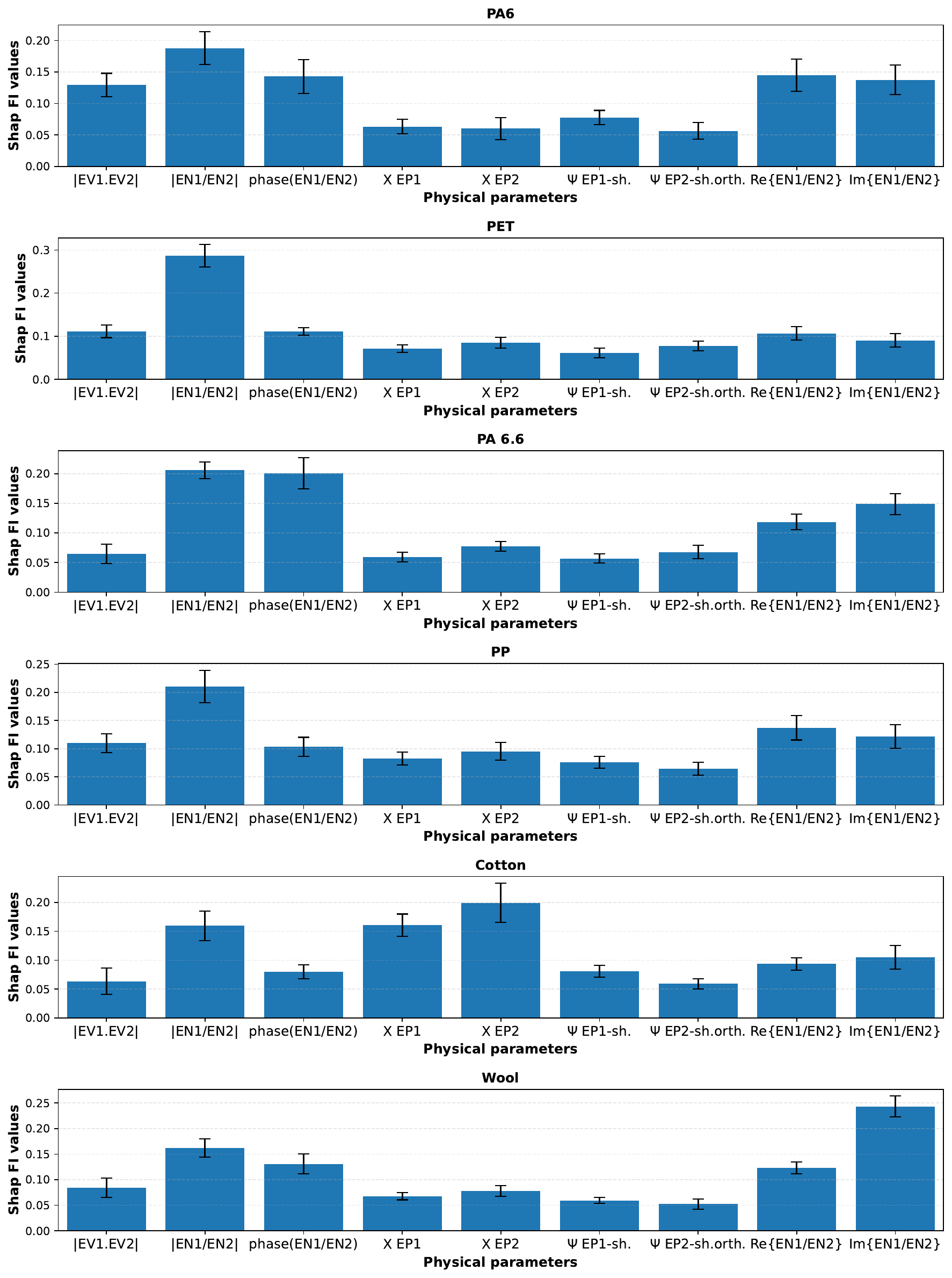}
  \caption{Class-wise SHAP-FI values of nine polarization characteristics. The most significant polarization characteristic for synthetic {MFFs} is the absolute value of the EN ratio, i.e. $\left|EN_{1}/EN_{2}\right|$. For natural {MFFs}, $\chi_{EP2}$ appears to be the most important indicator for cotton and Im$(EN_{1}/EN_{2})$ for wool.}
  \label{fgr:Shap_class}
\end{figure}

In the previous paragraphs, we analyzed SHAP-FI values without distinguish individual material classes. For a deeper understanding, we provide a class-wise SHAP-FI value analysis. For clarity, we visualized the sum of SHAP-FI values of statistical parameters for individual polarization characteristics (see Fig.~\ref{fgr:Shap_class}). It turned out that the ratio $\left|EN_{1}/EN_{2}\right|$ is particularly significant for synthetic MFFs. For natural MFFs, the parameter $\chi_{EP2}$ is the most significant for cotton and the parameter Im$(EN_{1}/EN_{2})$ is the most significant for wool, respectively.

Above findings enable to correlate SHAP-FI analysis with most significant material features of MFFs.
Birefringence of MFFs originates from preferential alignment of molecular structures along the fiber, which leads to anisotropic polarizability and, consequently, to a refractive index difference parallel and perpendicular to the fiber. In particular, the intrinsic anisotropy of natural fibers appears due to the ordered arrangement of structural proteins or polysaccharides, i.e., keratin in the case of wool and crystalline cellulose in the case of cotton \cite{Nemr2012}.
Importantly, the significance of class-wise SHAP-FI values for EP ellipticities ($\chi_{EP1}$, $\chi_{EP2}$) of cotton compared to other classes (Fig.~\ref{fgr:Shap_class}) is the consequence of its twisted ribbon-like geometry, which arises due to the natural twisting of the cellulose-rich cell wall as it dries \cite{Nemr2012, Mondal2021}. Such a complex structure determines the spatial variation of cellulose microfibrils along the local shape of MFF, thus broadening the distributions of $\chi_{EP1}$ and $\chi_{EP2}$ compared to the remaining material classes \cite{Behal2022Toward}.

On the other hand, optical birefringence in polymer textile fibers originates mainly from the orientation of polymer chains, whose alignment is determined during the drawing process \cite{Haji2012}. 
%
In particular, PET exhibits the highest value of the so-called virtual birefringence among the considered synthetic materials, exceeding 0.22 \cite{Huijts1994, Okada2016}, which represents the theoretical maximum of birefringence achieved within ideally oriented material domains. This finding is consistent with our experimental observations that the relative absorption $\left|EN_{1}/EN_{2}\right|$ is the most significant for PET among the remaining material classes \cite{Behal2022Toward}, as a higher value of birefringence induces higher Fresnel losses compared to materials with an overall comparable mean value of refractive index but lower birefringence \cite{Lekner1991}. Indeed, the high importance of $\left|EN_{1}/EN_{2}\right|$ is also confirmed via the class-wise SHAP-FI values of PET (Fig.~\ref{fgr:Shap_class}), where it is relatively highest among all material classes, i.e. the SHAP-FI value of $\left|EN_{1}/EN_{2}\right|$ of PET is 2.6 times higher than the second most important SHAP-FI value of PET, i.e., phase($EN_{1}/EN_{2}$). In addition, a substantially high importance of $\left|EN_{1}/EN_{2}\right|$ is also apparent for PP via the analysis of SHAP-FI values (Fig.~\ref{fgr:Shap_class}).

Finally, both the overall and the class-wise SHAP-FI analysis identifies the EN-based features as the most important, as can be noticed in Fig.~\ref{fgr:Shap} and Fig.~\ref{fgr:Shap_class}, respectively. To confirm the significance of $\left|EN_{1}/EN_{2}\right|$ and phase($EN_{1}/EN_{2}$) we selected all the relevant features (marked as 9-24 in Fig.~\ref{fgr:Shap}) and created an additional {fully-connected} neural-network based model without any changes to the previous architecture, {except number of neurons in the input layer, i.e., 16 neurons.} The model thus enables classification using these 16 features related to ENs. 
Indeed, the final classification accuracy evaluated on validation dataset reached 93.3~\%. This means that although the accuracy decreases by approximately 3.4~\% compared to the previous model, it still outperforms the classification performance of any common classifier exploiting all 72 features or even when applying feature selection (Table~\ref{tbl:class_compare}). 
The proposed SHAP-FI analysis thus identified the most significant features based on ENs which allow efficient MFF classification.
{Furthermore, we used the selected 16-feature dataset for classification by other classifiers to assess whether the reduced feature set improves the performance of lighter models. With the exception of Random Forest classifier, all models exhibited reduced performance compared to the results in Table~\ref{tbl:class_compare}. The only improvement was observed only for the Random Forest model, which achieved an accuracy of 0.88, stating that none of these classifiers outperformed our fully-connected neural-network based models.}

\section{Conclusions}
In this manuscript, we have proposed a deep-neural network–based approach for the classification of MFFs. Initially, a polarization-sensitive digital holographic microscope was used to measure polarization characteristics of MFFs, which were extracted from the reconstructed Jones matrix, including EPs defined by their orientation ($\psi$) and ellipticity ($\chi$) angles, and associated ENs describing amplitude and phase modulation. From these, nine polarization characteristics were derived, describing polarization homogeneity, modulation, ellipticity, and EP orientations related to the local fiber geometry. For each polarization characteristic, eight statistical parameters (mean, median, mode, mean absolute deviation, median absolute deviation, standard deviation, skewness, and kurtosis) were computed across all relevant pixels, yielding 72 features per MFF. In total, 296 MFFs from six material classes (PA6, PET, PP, PA6.6, cotton, and wool) were analyzed to form the input dataset for subsequent classification.

A classification method using a {fully-connected} neural network was implemented to classify MFF types from the fixed-length feature vectors. The architecture comprised of an input layer with 72 elements, {which corresponds to the number of features}, followed by four hidden layers with progressively decreasing sizes to extract increasingly abstract representations. Batch normalization and dropout were incorporated to reduce overfitting, and class weights were used to address dataset imbalance. The output layer generated probabilities for six classes. The model was trained using the Adam optimizer and categorical cross-entropy loss, with performance monitored via categorical accuracy. The proposed neural network classifier achieved high accuracy in distinguishing the considered MFF types. The accuracy on the validation dataset reached 96.7~\%, while on the complete dataset it achieved 98.6~\% accuracy.

Comparative evaluation against other machine learning methods, including Gaussian naive Bayes, random forest, support vector machine, K-nearest neighbors, gradient boosting, and logistic regression, confirmed the superior performance of the {fully-connected} deep neural network, as reflected by consistently higher accuracy, precision, recall, and F1 score on the validation dataset. In particular, logistic regression performed best among the conventional methods, achieving 92~\% classification accuracy, while the accuracy of the remaining methods ranged from 82~\% to 90~\%. The proposed deep neural network based approach thus improves the classification accuracy of at least 5~\% compared to the well-established statistical and machine learning classification approaches.

Furthermore, we quantified significance of the used polarization-derived features via SHAP-FI values. As a result, the most significant polarization characteristic is the absolute value of ENs ratio, i.e. $(\left| EN_{1}/EN_{2}\right|)$, with its statistical parameters—particularly the median and median absolute deviation—emerging as the strongest contributors among all features. Overall, the EN-based characteristics dominate the classification process, collectively surpassing the contributions of EV- and EP-based features.

Moreover, additional class-wise SHAP-FI analysis predicted the significant influence of EN based features, with exception of cotton, where ellipticity $\chi_{EP2}$ proved to be the most important. The significance of ENs was confirmed via additional {fully-connected} neural-network based classification model with the preserved architecture, while exploiting features relevant just to polarization characteristics $\left|EN_{1}/EN_{2}\right|$ and phase($EN_{1}/EN_{2}$). The final classification accuracy reached 93.3~\%, thus still outperforming the classification performance of any above-noted machine-learning classifiers (see Table \ref{tbl:class_compare}). {Nevertheless, reducing the number of input parameters is generally not desirable when aiming for the ultimate classification performance.}

{Moreover, comparison of the achieved accuracies of both trained fully-connected deep neural network models provides important information. Especially, the assumption that small datasets with many monitored parameters inevitably lead to redundancy or overfitting is not always justified. In fact, a more sophisticated model, capable of capturing the underlying nonlinear relationships within the data, can improve predictive accuracy because of its stronger expressivity and generalization. Indeed, the high classification accuracy on the validation dataset achieved with the 72-feature model (96.7\%) and the reduced 16-feature model (93.3\%) suggest that the deep neural network captures meaningful underlying structure rather than relying solely on redundant inputs, while the slight decrease of the performance (from 96.7\% to 93.3\%) indicates that additional features contribute complementary information. This result also approves the use of the full set of 72 features for the optimal classification performance.}

In overall, the achieved results show that polarization-derived features contain significant information for classifying MFFs. When combined with deep learning, these features enable highly accurate and interpretable classification, demonstrating the potential of polarization-sensitive digital holography as an effective tool for microplastic detection and analysis. Based on the previous clarifications, we retain our novelty claims. Specifically, we demonstrate for the first time that deep learning with a large, overparameterized network can be successfully applied to small real-world microfiber datasets, achieving performance that surpasses the current state of the art. We further present deep-neural-network-based MFF classification relying exclusively on polarization-based features, with additional performance gains expected through the incorporation of complementary feature types, such as fractal, DHM-based, and Haralick texture features. Finally, we provide the first explainability analysis of a deep learning model in this application.

{Note added. During the review process, we became aware of independent related work by Yang et al. \cite{Yang2026}, which develops an approach closely related to the one presented in this article.}

\section*{Funding}
J.B. acknowledges project of the Czech Science Foundation (GAČR), project No. 25-17712I. J.A. acknowledges projects IGA\_PrF\_2025\_005 and IGA\_PrF\_2026\_003 of Palacký University Olomouc.
This work was partially supported by project ROMEO-smaRt Online Multisensory systEm for microplastic quantificatiOn and water quality assessment (project CUP: B83C24009260005).

\section*{Acknowledgments}
We acknowledge the use of cluster computing resources provided by the Department of Optics, Palacký University Olomouc. We thank J. Provazník for maintaining the cluster and providing support.

\section*{Disclosures}
The authors declare no conflicts of interest.

\section*{Data Availability}
Data are publicly available at \cite{repository}

\bibliography{reference_library}

@article{Gesamp2020Sources,
author = {Koehler, Angela and Anderson, Alison and Andrady, Anthony and Arthur, Courtney and Bouwman, Hindrik and Gall, Sarah and Hidalgo-Ruz, Valeria and Koehler, Angela and Law, Kara and Leslie, Heather and Kershaw, Peter and Pahl, Sabine and Potemra, Jim and Ryan, Peter and Shim, Won and Thompson, Richard and Takada, Hideshige and Turra, Alexander and Wyles, Kayleigh},
year = {2015},
month = {04},
volume = {90},
pages = {96},
title = {Sources, fate and effects of microplastics in the marine environment: a global assessment},
doi = {10.13140/RG.2.1.3803.7925},
journal = {Journal Series GESAMP Reports and Studies}
}

@article{Mazhandu2020Integrated,
  title = {Integrated and Consolidated Review of Plastic Waste Management and Bio-Based Biodegradable Plastics: Challenges and Opportunities}, 
  author = {Mazhandu, Zvanaka S. and Muzenda, Edison and Mamvura, Tirivaviri A. and Belaid, Mohamed and Nhubu, Trust}, 
  year = {2020}, 
  month = {10}, 
  day = {12}, 
  number = {20}, 
  volume = {12}, 
  journal = {Sustainability}, 
  pages = {8360}, 
  publisher = {MDPI AG}, 
  url = {http://dx.doi.org/10.3390/su12208360}, 
  doi = {10.3390/su12208360}
}

@article{Sun2019Microplastics,
  title = {Microplastics in wastewater treatment plants: Detection, occurrence and removal}, 
  author = {Sun, Jing and Dai, Xiaohu and Wang, Qilin and van Loosdrecht, Mark C.M. and Ni, Bing-Jie}, 
  year = {2019}, 
  volume = {152}, 
  journal = {Water Research}, 
  pages = {21-37}, 
  publisher = {Elsevier BV}, 
  url = {http://dx.doi.org/10.1016/j.watres.2018.12.050}, 
  doi = {10.1016/j.watres.2018.12.050}
}

@article{Henry2019Microfibres,
  title = {Microfibres from apparel and home textiles: Prospects for including microplastics in environmental sustainability assessment}, 
  author = {Henry, Beverley and Laitala, Kirsi and Klepp, Ingun Grimstad}, 
  year = {2019}, 
  volume = {652}, 
  journal = {Science of The Total Environment}, 
  pages = {483-494}, 
  publisher = {Elsevier BV}, 
  url = {http://dx.doi.org/10.1016/j.scitotenv.2018.10.166}, 
  doi = {10.1016/j.scitotenv.2018.10.166}
}

@inbook{Boucher2017Primary,
author = {Boucher, Julien and Friot, Damien},
year = {2017},
month = {01},
title = {Primary Microplastics in the Oceans: A Global Evaluation of Sources},
pages = {20--22},
url = {https://doi.org/10.2305/iucn.ch.2017.01.en},
isbn = {978-2-8317-1827-9},
doi = {10.2305/IUCN.CH.2017.01.en},
publisher = {IUCN}
}

@article{Shim2017Identification,
  title = {Identification methods in microplastic analysis: a review}, 
  author = {Shim, Won Joon and Hong, Sang Hee and Eo, Soeun Eo}, 
  year = {2017}, 
  number = {9}, 
  volume = {9}, 
  journal = {Analytical Methods}, 
  pages = {1384-1391}, 
  publisher = {Royal Society of Chemistry (RSC)}, 
  url = {http://dx.doi.org/10.1039/c6ay02558g}, 
  doi = {10.1039/c6ay02558g}
}

@article{Hildebrandt2019Evaluation,
  title = {Evaluation of continuous flow centrifugation as an alternative technique to sample microplastic from water bodies}, 
  author = {Hildebrandt, L. and Voigt, N. and Zimmermann, T. and Reese, A. and Proefrock, D.}, 
  year = {2019}, 
  volume = {151}, 
  journal = {Marine Environmental Research}, 
  pages = {104768}, 
  publisher = {Elsevier BV}, 
  url = {http://dx.doi.org/10.1016/j.marenvres.2019.104768}, 
  doi = {10.1016/j.marenvres.2019.104768}
}

@article{DeFalco2018Evaluation,
title = {Evaluation of microplastic release caused by textile washing processes of synthetic fabrics},
journal = {Environmental Pollution},
volume = {236},
pages = {916-925},
year = {2018},
issn = {0269-7491},
doi = {https://doi.org/10.1016/j.envpol.2017.10.057},
url = {https://www.sciencedirect.com/science/article/pii/S0269749117309387},
author = {Francesca {De Falco} and Maria Pia Gullo and Gennaro Gentile and Emilia {Di Pace} and Mariacristina Cocca and Laura Gelabert and Marolda Brouta-Agnésa and Angels Rovira and Rosa Escudero and Raquel Villalba and Raffaella Mossotti and Alessio Montarsolo and Sara Gavignano and Claudio Tonin and Maurizio Avella},
}

@article{Maes2017A,
  title = {A rapid-screening approach to detect and quantify microplastics based on fluorescent tagging with Nile Red}, 
  author = {Maes, Thomas and Jessop, Rebecca and Wellner, Nikolaus and Haupt, Karsten and Mayes, Andrew G.}, 
  year = {2017}, 
  month = {3}, 
  day = {16}, 
  number = {1}, 
  volume = {7}, 
  journal = {Scientific Reports}, 
  pages = {44501}, 
  publisher = {Springer Science and Business Media LLC}, 
  url = {http://dx.doi.org/10.1038/srep44501}, 
  doi = {10.1038/srep44501}
}

@article{Mintenig2017Identification,
  title = {Identification of microplastic in effluents of waste water treatment plants using focal plane array-based micro-Fourier-transform infrared imaging}, 
  author = {Mintenig, S.M. and Int-Veen, I. and L{\"o}der, M.G.J. and Primpke, S. and Gerdts, G.}, 
  year = {2017}, 
  volume = {108}, 
  journal = {Water Research}, 
  pages = {365-372}, 
  publisher = {Elsevier BV}, 
  url = {http://dx.doi.org/10.1016/j.watres.2016.11.015}, 
  doi = {10.1016/j.watres.2016.11.015}
}

@Article{Primpke2017An,
author ="Primpke, S. and Lorenz, C. and Rascher-Friesenhausen, R. and Gerdts, G.",
title  ="An automated approach for microplastics analysis using focal plane array (FPA) FTIR microscopy and image analysis",
journal  ="Anal. Methods",
year  ="2017",
volume  ="9",
issue  ="9",
pages  ="1499-1511",
publisher  ="The Royal Society of Chemistry",
doi  ="10.1039/C6AY02476A",
url  ="http://dx.doi.org/10.1039/C6AY02476A",
}

@article{Circelli2024Comparison,
title = {Comparison of ATR-FTIR and NIR spectroscopy for identification of microplastics in biosolids},
journal = {Science of The Total Environment},
volume = {916},
pages = {170215},
year = {2024},
issn = {0048-9697},
doi = {https://doi.org/10.1016/j.scitotenv.2024.170215},
url = {https://www.sciencedirect.com/science/article/pii/S0048969724003504},
author = {Luana Circelli and Zhongqi Cheng and Evan Garwood and Kerem Yuksel and Erika {Di Iorio} and Ruggero Angelico and Claudio Colombo}
}

@article{Zhang2018Identification,
author = {Zhang, Jixiong and Tian, Kuangda and Lei, Chunli and Min, Shungeng},
year = {2018},
month = {04},
pages = {2881–2887},
title = {Identification and quantification of microplastics in table sea salts using micro-NIR imaging methods},
volume = {10},
journal = {Analytical Methods},
doi = {10.1039/C8AY00125A}
}

@article{Seghers2021Preparation,
  title = {Preparation of a reference material for microplastics in water{---}evaluation of homogeneity}, 
  author = {Seghers, John and Stefaniak, Elzbieta A. and La Spina, Rita and Cella, Claudia and Mehn, Dora and Gilliland, Douglas and Held, Andrea and Jacobsson, Ulf and Emteborg, H{\aa}kan}, 
  year = {2021}, 
  month = {2}, 
  day = {6}, 
  number = {1}, 
  volume = {414}, 
  journal = {Analytical and Bioanalytical Chemistry}, 
  pages = {385-397}, 
  publisher = {Springer Science and Business Media LLC}, 
  url = {http://dx.doi.org/10.1007/s00216-021-03198-7}, 
  doi = {10.1007/s00216-021-03198-7}
}

@article{Cabernard2018Comparison,
author = {Cabernard, Livia and Roscher, Lisa and Lorenz, Claudia and Gerdts, Gunnar and Primpke, Sebastian},
title = {Comparison of Raman and Fourier Transform Infrared Spectroscopy for the Quantification of Microplastics in the Aquatic Environment},
journal = {Environmental Science \& Technology},
volume = {52},
number = {22},
pages = {13279-13288},
year = {2018},
doi = {10.1021/acs.est.8b03438},
URL = {https://doi.org/10.1021/acs.est.8b03438}
}

@article{Zarfl2019Promising,
author = {Zarfl, Christiane},
year = {2019},
month = {03},
pages = {3743–3756},
title = {Promising techniques and open challenges for microplastic identification and quantification in environmental matrices},
volume = {411},
journal = {Analytical and Bioanalytical Chemistry},
doi = {10.1007/s00216-019-01763-9}
}

@article{Bianco2020Microplastic,
  title = {Microplastic Identification via Holographic Imaging and Machine Learning}, 
  author = {Bianco, Vittorio and Memmolo, Pasquale and Carcagn{\`\i}, Pierluigi and Merola, Francesco and Paturzo, Melania and Distante, Cosimo and Ferraro, Pietro}, 
  year = {2020}, 
  month = {1}, 
  day = {13}, 
  number = {2}, 
  volume = {2}, 
  journal = {Advanced Intelligent Systems}, 
  pages = {1900153}, 
  publisher = {Wiley}, 
  url = {http://dx.doi.org/10.1002/aisy.201900153}, 
  doi = {10.1002/aisy.201900153}
}

@article{Nayak2021A,
author = {Nayak, Aditya and Malkiel, Ed and McFarland, Malcolm and Twardowski, Michael and Sullivan, James},
year = {2021},
month = {01},
pages = {572147},
title = {A Review of Holography in the Aquatic Sciences: In situ Characterization of Particles, Plankton, and Small Scale Biophysical Interactions},
volume = {7},
journal = {Frontiers in Marine Science},
doi = {10.3389/fmars.2020.572147}
}

@article{Takahashi2020Identification,
author = {Tomoko Takahashi and Zonghua Liu and Thangavel Thevar and Nicholas Burns and Sumeet Mahajan and Dhugal Lindsay and John Watson and Blair Thornton},
journal = {Appl. Opt.},
number = {17},
pages = {5073--5078},
publisher = {Optica Publishing Group},
title = {Identification of microplastics in a large water volume by integrated holography and Raman spectroscopy},
volume = {59},
month = {Jun},
year = {2020},
url = {https://opg.optica.org/ao/abstract.cfm?URI=ao-59-17-5073},
doi = {10.1364/AO.393643}
}

@article{Zhu2021Microplastic,
doi = {10.1088/2515-7647/abf250},
url = {https://doi.org/10.1088/2515-7647/abf250},
year = {2021},
month = {apr},
publisher = {IOP Publishing},
volume = {3},
number = {2},
pages = {024013},
author = {Zhu, Yanmin and Yeung, Chok Hang and Lam, Edmund Y},
title = {Microplastic pollution monitoring with holographic classification and deep learning},
journal = {Journal of Physics: Photonics}
}

@article{Russo2024Deep,
  title = {Deep Classification of Microplastics Through Image Fusion Techniques}, 
  author = {Russo, Paolo and Di Ciaccio, Fabiana}, 
  year = {2024}, 
  volume = {12}, 
  journal = {IEEE Access}, 
  pages = {134852-134861}, 
  publisher = {Institute of Electrical and Electronics Engineers (IEEE)}, 
  url = {http://dx.doi.org/10.1109/access.2024.3423661}, 
  doi = {10.1109/access.2024.3423661}
}

@article{Bianco2021Identification,
author = {Bianco, Vittorio and Pirone, Daniele and Memmolo, P. and Merola, Francesco and Ferraro, Pietro},
year = {2021},
month = {06},
pages = {2148-2157},
title = {Identification of Microplastics Based on the Fractal Properties of Their Holographic Fingerprint},
volume = {8},
journal = {ACS Photonics},
doi = {10.1021/acsphotonics.1c00591}
}

@article{Owen2021Microplastic,
  title = {Microplastic adulteration in homogenized fish and seafood - a mid-infrared and machine learning proof of concept}, 
  author = {Owen, Stephanie and Cureton, Samuel and Szuhan, Mathew and McCarten, Joel and Arvanitis, Panagiota and Ascione, Max and Truong, Vi Khanh and Chapman, James and Cozzolino, Daniel}, 
  year = {2021}, 
  volume = {260}, 
  journal = {Spectrochimica Acta Part A: Molecular and Biomolecular Spectroscopy}, 
  pages = {119985}, 
  publisher = {Elsevier BV}, 
  url = {http://dx.doi.org/10.1016/j.saa.2021.119985}, 
  doi = {10.1016/j.saa.2021.119985}
}

@article{DaSilva2020Classification,
author = {da Silva, Vitor H. and Murphy, Fionn and Amigo, Jos{\'e} M. and Stedmon, Colin and Strand, Jakob},
title = {Classification and Quantification of Microplastics (<100 $\mu$m) Using a Focal Plane Array–Fourier Transform Infrared Imaging System and Machine Learning},
journal = {Analytical Chemistry},
volume = {92},
number = {20},
pages = {13724-13733},
year = {2020},
doi = {10.1021/acs.analchem.0c01324},
URL = {https://doi.org/10.1021/acs.analchem.0c01324}
}

@article{Kedzierski2019A,
title = {A machine learning algorithm for high throughput identification of FTIR spectra: Application on microplastics collected in the Mediterranean Sea},
journal = {Chemosphere},
volume = {234},
pages = {242-251},
year = {2019},
issn = {0045-6535},
doi = {https://doi.org/10.1016/j.chemosphere.2019.05.113},
url = {https://www.sciencedirect.com/science/article/pii/S0045653519310197},
author = {Mikaël Kedzierski and Mathilde Falcou-Préfol and Marie Emmanuelle Kerros and Maryvonne Henry and Maria Luiza Pedrotti and Stéphane Bruzaud}
}

@article{Ballard2021Machine,
author = {Ballard, Zachary and Brown, Calvin and Madni, Asad and Ozcan, Aydogan},
year = {2021},
month = {06},
pages = {1-10},
title = {Machine learning and computation-enabled intelligent sensor design},
volume = {3},
journal = {Nature Machine Intelligence},
doi = {10.1038/s42256-021-00360-9}
}

@article{Yuan2023Geometric,
  title = {Geometric deep optical sensing}, 
  author = {Yuan, Shaofan and Ma, Chao and Fetaya, Ethan and Mueller, Thomas and Naveh, Doron and Zhang, Fan and Xia, Fengnian}, 
  year = {2023}, 
  month = {3}, 
  day = {17}, 
  number = {6637}, 
  volume = {379}, 
  journal = {Science}, 
  pages = {eade1220}, 
  publisher = {American Association for the Advancement of Science (AAAS)}, 
  url = {http://dx.doi.org/10.1126/science.ade1220}, 
  doi = {10.1126/science.ade1220}
}

@article{Freire2023Artificial,
author = {Freire, Pedro and Manuylovich, Egor and Prilepsky, Jaroslaw and Turitsyn, S.k},
year = {2023},
month = {09},
pages = {739-834},
title = {Artificial neural networks for photonic applications—from algorithms to implementation: tutorial},
volume = {15},
journal = {Advances in Optics and Photonics},
doi = {10.1364/AOP.484119}
}

@article{Bommasani2021On,
  author       = {Rishi Bommasani and
                  Drew A. Hudson and
                  Ehsan Adeli and
                  Russ B. Altman and
                  Simran Arora and
                  Sydney von Arx and
                  Michael S. Bernstein and
                  Jeannette Bohg and
                  Antoine Bosselut and
                  Emma Brunskill and
                  Erik Brynjolfsson and
                  Shyamal Buch and
                  Dallas Card and
                  Rodrigo Castellon and
                  Niladri S. Chatterji and
                  Annie S. Chen and
                  Kathleen Creel and
                  Jared Quincy Davis and
                  Dorottya Demszky and
                  Chris Donahue and
                  Moussa Doumbouya and
                  Esin Durmus and
                  Stefano Ermon and
                  John Etchemendy and
                  Kawin Ethayarajh and
                  Li Fei{-}Fei and
                  Chelsea Finn and
                  Trevor Gale and
                  Lauren E. Gillespie and
                  Karan Goel and
                  Noah D. Goodman and
                  Shelby Grossman and
                  Neel Guha and
                  Tatsunori Hashimoto and
                  Peter Henderson and
                  John Hewitt and
                  Daniel E. Ho and
                  Jenny Hong and
                  Kyle Hsu and
                  Jing Huang and
                  Thomas Icard and
                  Saahil Jain and
                  Dan Jurafsky and
                  Pratyusha Kalluri and
                  Siddharth Karamcheti and
                  Geoff Keeling and
                  Fereshte Khani and
                  Omar Khattab and
                  Pang Wei Koh and
                  Mark S. Krass and
                  Ranjay Krishna and
                  Rohith Kuditipudi and
                  et al.},
  title        = {On the Opportunities and Risks of Foundation Models},
  journal      = {CoRR},
  volume       = {abs/2108.07258},
  year         = {2021},
  url          = {https://arxiv.org/abs/2108.07258},
  eprinttype    = {arXiv},
  eprint       = {2108.07258},
  bibsource    = {dblp computer science bibliography, https://dblp.org}
}

@article{Sierra2020Identification,
author = {Sierra, Ignacio and Rodriguez, Mauricio and Faccio, Ricardo and Carrizo, Daniel and Fornaro, Laura and Pérez Parada, Andrés},
year = {2020},
month = {03},
pages = {7409–7419},
title = {Identification of microplastics in wastewater samples by means of polarized light optical microscopy},
volume = {27},
journal = {Environmental Science and Pollution Research},
doi = {10.1007/s11356-019-07011-y}
}

@article{Wang2008Jones,
  title = {Jones phase microscopy of transparent and anisotropic samples}, 
  author = {Wang, Zhuo and Millet, Larry J. and Gillette, Martha U. and Popescu, Gabriel}, 
  year = {2008}, 
  month = {5}, 
  day = {30}, 
  number = {11}, 
  volume = {33}, 
  journal = {Optics Letters}, 
  pages = {1270}, 
  publisher = {Optica Publishing Group}, 
  url = {http://dx.doi.org/10.1364/ol.33.001270}, 
  doi = {10.1364/ol.33.001270}
}

@article{Jiao2020Real,
author = {Jiao, Yuheng and Kandel, Mikhail and Liu, Xiaojun and Lu, Wenlong and Popescu, Gabriel},
year = {2020},
month = {10},
pages = {34190-34200},
title = {Real-time Jones phase microscopy for studying transparent and birefringent specimens},
volume = {28},
journal = {Optics Express},
doi = {10.1364/OE.397062}
}

@article{Behal2022Toward,
author = {Běhal, Jaromír and Valentino, Marika and Miccio, Lisa and Bianco, Vittorio and Itri, Simona and Mossotti, Raffaella and Dalla Fontana, Giulia and Stella, Ettore and Ferraro, Pietro},
year = {2022},
month = {01},
pages = {694-705},
title = {Toward an All-Optical Fingerprint of Synthetic and Natural Microplastic Fibers by Polarization-Sensitive Holographic Microscopy},
volume = {9},
journal = {ACS Photonics},
doi = {10.1021/acsphotonics.1c01781}
}

@article{Valentino2022Intelligent,
  title = {Intelligent polarization-sensitive holographic flow-cytometer: Towards specificity in classifying natural and microplastic fibers}, 
  author = {Valentino, Marika and B{\u{e}}hal, Jarom{\'\i}r and Bianco, Vittorio and Itri, Simona and Mossotti, Raffaella and Fontana, Giulia Dalla and Battistini, Tiziano and Stella, Ettore and Miccio, Lisa and Ferraro, Pietro}, 
  year = {2022}, 
  volume = {815}, 
  journal = {Science of The Total Environment}, 
  pages = {152708}, 
  publisher = {Elsevier BV}, 
  url = {http://dx.doi.org/10.1016/j.scitotenv.2021.152708}, 
  doi = {10.1016/j.scitotenv.2021.152708}
}

@article{Valentino2024Discern,
author = {Valentino, Marika and Běhal, Jaromír and Tonetti, C. and Carletto, R.A. and Itri, Simona and Memmolo, P. and Stella, Ettore and Miccio, L. and Bianco, Vittorio and Ferraro, Pietro},
year = {2024},
month = {10},
pages = {108395},
title = {Discernment of textile fibers by polarization-sensitive Digital Holographic microscope and machine learning},
volume = {181},
journal = {Optics and Lasers in Engineering},
doi = {10.1016/j.optlaseng.2024.108395}
}

@article{Lu1994Homogeneous,
  title = {Homogeneous and inhomogeneous Jones matrices}, 
  author = {Lu, Shih-Yau and Chipman, Russell A.}, 
  year = {1994}, 
  month = {2}, 
  day = {1}, 
  number = {2}, 
  volume = {11}, 
  journal = {Journal of the Optical Society of America A}, 
  pages = {766}, 
  publisher = {Optica Publishing Group}, 
  url = {http://dx.doi.org/10.1364/josaa.11.000766}, 
  doi = {10.1364/josaa.11.000766}
}

@article{Baroni2020Extending,
  title = {Extending Quantitative Phase Imaging to Polarization-Sensitive Materials}, 
  author = {Baroni, Arthur and Chamard, Virginie and Ferrand, Patrick}, 
  year = {2020}, 
  month = {5}, 
  day = {12}, 
  number = {5}, 
  volume = {13}, 
  journal = {Physical Review Applied}, 
  pages = {054028}, 
  publisher = {American Physical Society (APS)}, 
  url = {http://dx.doi.org/10.1103/physrevapplied.13.054028}, 
  doi = {10.1103/physrevapplied.13.054028}
}

@article{MLComparArticle,
title = {A Comparative Analysis of Machine Learning Algorithms for Classification Purpose},
journal = {Procedia Computer Science},
volume = {215},
pages = {422-431},
year = {2022},
note = {4th International Conference on Innovative Data Communication Technology and Application},
issn = {1877-0509},
doi = {https://doi.org/10.1016/j.procs.2022.12.044},
url = {https://www.sciencedirect.com/science/article/pii/S1877050922021159},
author = {Vraj Sheth and Urvashi Tripathi and Ankit Sharma}
}

@article{MLResearch_article,
author = {Sarker, Iqbal},
year = {2021},
month = {03},
pages = {160},
title = {Machine Learning: Algorithms, Real-World Applications and Research Directions},
volume = {2},
journal = {SN Computer Science},
doi = {10.1007/s42979-021-00592-x}
}

@article{Liu2025Recent,
author = {Liu, Jinhui and Niu, Jiaqi and Wu, Wanqing and Zhang, Ziyang and Ning, Ye and Zheng, Qinggong},
year = {2025},
month = {02},
pages = {117695},
title = {Recent advances in the detection of microplastics in the aqueous environment by electrochemical sensors: A review},
volume = {214},
journal = {Marine pollution bulletin},
doi = {10.1016/j.marpolbul.2025.117695}
}

@article{Lukose2025Gaining,
  title = {Gaining traction of optical modalities in the detection of microplastics}, 
  author = {Lukose, Jijo and Sunil, Megha and Westhead, Elizabeth K and Chidangil, Santhosh and Kumar, Satheesh}, 
  year = {2025}, 
  volume = {47}, 
  journal = {Current Opinion in Chemical Engineering}, 
  pages = {101086}, 
  publisher = {Elsevier BV}, 
  url = {http://dx.doi.org/10.1016/j.coche.2024.101086}, 
  doi = {10.1016/j.coche.2024.101086}
}

@article{Yan2024Pushing,
  title = {Pushing the frontiers of micro/nano-plastic detection with portable instruments}, 
  author = {Yan, Yuhao and Zeng, Li and Gao, Jie and Cheng, Jiexia and Zheng, Xuehan and Wang, Guangxuan and Ding, Yun and Zhao, Jing and Qin, Hua and Zhao, Chao and Luo, Qian and Liu, Runzeng and Chen, Liqun and Cai, Zongwei and Yan, Bing and Qu, Guangbo and Jiang, Guibin}, 
  year = {2024}, 
  volume = {181}, 
  journal = {TrAC Trends in Analytical Chemistry}, 
  pages = {118044}, 
  publisher = {Elsevier BV}, 
  url = {http://dx.doi.org/10.1016/j.trac.2024.118044}, 
  doi = {10.1016/j.trac.2024.118044}
}

@article{Jianqing2023Snapshot,
author = {Huang, Jianqing and Zhu, Yanmin and Li, Yuxing and Lam, Edmund},
year = {2023},
month = {11},
pages = {4483-4493},
title = {Snapshot Polarization-Sensitive Holography for Detecting Microplastics in Turbid Water},
volume = {10},
journal = {ACS Photonics},
doi = {10.1021/acsphotonics.3c01350}
}

@article{Yoganandham2023Micro,
  title = {Micro(nano)plastics in commercial foods: A review of their characterization and potential hazards to human health}, 
  author = {Yoganandham, Suman Thodhal and Hamid, Naima and Junaid, Muhammad and Duan, Jin-Jing and Pei, De-Sheng}, 
  year = {2023}, 
  volume = {236}, 
  journal = {Environmental Research}, 
  pages = {116858}, 
  publisher = {Elsevier BV}, 
  url = {http://dx.doi.org/10.1016/j.envres.2023.116858}, 
  doi = {10.1016/j.envres.2023.116858}
}

@article{Li2023Recognition,
  title = {Recognition of microplastics suspended in seawater via refractive index by Mueller matrix polarimetry}, 
  author = {Li, Jiajin and Liu, Hongyuan and Liao, Ran and Wang, Hongjian and Chen, Yan and Xiang, Jing and Xu, Xiangrong and Ma, Hui}, 
  year = {2023}, 
  volume = {188}, 
  journal = {Marine Pollution Bulletin}, 
  pages = {114706}, 
  publisher = {Elsevier BV}, 
  url = {http://dx.doi.org/10.1016/j.marpolbul.2023.114706}, 
  doi = {10.1016/j.marpolbul.2023.114706}
}

@inproceedings{Lundberg2017,
  title={A Unified Approach to Interpreting Model Predictions},
  author={Scott M. Lundberg and Su-In Lee},
  booktitle={Neural Information Processing Systems},
  year={2017},
  url={https://api.semanticscholar.org/CorpusID:21889700},
  journal = {adv. Neural Inf. Process. Sys.},
  volume = {20},
  pages = {4765-4774},
}

@article{Lundberg2020,
  title = {From local explanations to global understanding with explainable AI for trees}, 
  author = {Lundberg, Scott M. and Erion, Gabriel and Chen, Hugh and DeGrave, Alex and Prutkin, Jordan M. and Nair, Bala and Katz, Ronit and Himmelfarb, Jonathan and Bansal, Nisha and Lee, Su-In}, 
  year = {2020}, 
  month = {1}, 
  day = {17}, 
  number = {1}, 
  volume = {2}, 
  journal = {Nature Machine Intelligence}, 
  pages = {56-67}, 
  publisher = {Springer Science and Business Media LLC}, 
  url = {http://dx.doi.org/10.1038/s42256-019-0138-9}, 
  doi = {10.1038/s42256-019-0138-9}
}

@article{PonceBobadilla2024,
author = {Ponce Bobadilla, Ana Victoria and Schmitt, Vanessa and Maier, Corinna and Mensing, Sven and Stodtmann, Sven},
year = {2024},
month = {10},
pages = {},
title = {Practical guide to SHAP analysis: Explaining supervised machine learning model predictions in drug development},
volume = {17},
journal = {Clinical and Translational Science},
doi = {10.1111/cts.70056}
}

@book{Goodfellow2016Deep,
    title={Deep Learning},
    author={Ian Goodfellow and Yoshua Bengio and Aaron Courville},
    publisher={MIT Press},
    note={\url{http://www.deeplearningbook.org}},
    year={2016}
}

@article{Srivastava2014Dropout,
author = {Srivastava, Nitish and Hinton, Geoffrey and Krizhevsky, Alex and Sutskever, Ilya and Salakhutdinov, Ruslan},
year = {2014},
month = {06},
pages = {1929-1958},
title = {Dropout: A Simple Way to Prevent Neural Networks from Overfitting},
volume = {15},
journal = {Journal of Machine Learning Research}
}

@article{Kingma2015A,
  title={Adam: A Method for Stochastic Optimization},
  author={Diederik P. Kingma and Jimmy Ba},
  journal={CoRR},
  year={2014},
  volume={abs/1412.6980},
  url={https://api.semanticscholar.org/CorpusID:6628106}
}

@article{Lee2023Automatic,
  title = {Automatic classification of microplastics and natural organic matter mixtures using a deep learning moidel}, 
  author = {Lee, Seunghyeon and Jeong, Heewon and Hong, Seok Min and Yun, Daeun and Lee, Jiye and Kim, Eunju and Cho, Kyung Hwa}, 
  year = {2023}, 
  volume = {246}, 
  journal = {Water Research}, 
  pages = {120710}, 
  publisher = {Elsevier BV}, 
  url = {http://dx.doi.org/10.1016/j.watres.2023.120710}, 
  doi = {10.1016/j.watres.2023.120710}
}

@article{Zhang2023A,
author = {Zhang, Wei and Feng, Weiwei and Cai, Zongqi and Wang, Huanqing and Yan, Qi and Wang, Qing},
year = {2022},
month = {12},
pages = {103487},
title = {A deep one-dimensional convolutional neural network for microplastics classification using Raman spectroscopy},
volume = {124},
journal = {Vibrational Spectroscopy},
doi = {10.1016/j.vibspec.2022.103487}
}

@inproceedings{Gkioxari2017Mask,
  title = {Mask R-CNN},
  booktitle={2017 IEEE International Conference on Computer Vision (ICCV)},
  author = {He, Kaiming and Gkioxari, Georgia and Dollar, Piotr and Girshick, Ross}, 
  year = {2017}, 
  pages = {2961–2969},
  journal = {2017 IEEE International Conference on Computer Vision (ICCV)}, 
  publisher = {IEEE}, 
  url = {http://dx.doi.org/10.1109/iccv.2017.322}, 
  doi = {10.1109/iccv.2017.322}
}

@article{Han2023Deep,
  title = {Deep learning based approach for automated characterization of large marine microplastic particles}, 
  author = {Han, Xiao-Le and Jiang, Ning-Jun and Hata, Toshiro and Choi, Jongseong and Du, Yan-Jun and Wang, Yi-Jie}, 
  year = {2023}, 
  volume = {183}, 
  journal = {Marine Environmental Research}, 
  pages = {105829}, 
  publisher = {Elsevier BV}, 
  url = {http://dx.doi.org/10.1016/j.marenvres.2022.105829}, 
  doi = {10.1016/j.marenvres.2022.105829}
}

@article{Akkajit2023Comparative,
  title = {Comparative analysis of five convolutional neural networks and transfer learning classification approach for microplastics in wastewater treatment plants}, 
  author = {Akkajit, Pensiri and Sukkuea, Arsanchai and Thongnonghin, Boonnisa}, 
  year = {2023}, 
  volume = {78}, 
  journal = {Ecological Informatics}, 
  pages = {102328}, 
  publisher = {Elsevier BV}, 
  url = {http://dx.doi.org/10.1016/j.ecoinf.2023.102328}, 
  doi = {10.1016/j.ecoinf.2023.102328}
}

@article{Cybenko1989,
author = {Lewicki, Grzegorz and Marino, Giuseppe},
year = {2004},
month = {12},
pages = {1147-1152},
title = {Approximation by Superpositions of a Sigmoidal Function},
volume = {17},
journal = {Appl. Math. Lett.},
doi = {10.1016/j.aml.2003.11.006}
}

@article{Lu2017,
author = {Lu, Zhou and Pu, Hongming and Wang, Feicheng and Hu, Zhiqiang and Wang, Liwei},
journal      = {CoRR},
year = {2017},
month = {09},
pages = {6232–6240},
title = {The Expressive Power of Neural Networks: A View from the Width},
doi = {10.48550/arXiv.1709.02540}
}

@inbook{Nemr2012,
  title     = "Textiles",
  editor    = "El-Nemr, Ahmed",
  publisher = "Nova Science",
  month     =  may,
  year      =  2012,
  isbn      = "978-0-12-821483-1",
  chapter = {2,4}
}

@ARTICLE{Haji2012,
  title     = "The effect of hot multistage drawing on molecular structure and
               optical properties of polyethylene terephthalate fibers",
  author    = "Haji, Aminoddin and Rahbar, Ruhollah Semnani and Kalantari,
               Bahareh",
  journal   = "Mater. Res.",
  publisher = "FapUNIFESP (SciELO)",
  volume    =  15,
  number    =  4,
  pages     = "554--560",
  month     =  jul,
  year      =  2012
}

@ARTICLE{Okada2016,
  title     = "Reliability of intrinsic birefringence estimated via the
               modified stress-optical rule",
  author    = "Okada, Yuki and Urakawa, Osamu and Inoue, Tadashi",
  journal   = "Polym. J.",
  publisher = "Springer Science and Business Media LLC",
  volume    =  48,
  number    =  11,
  pages     = "1073--1078",
  month     =  nov,
  year      =  2016,
  language  = "en"
}

@ARTICLE{Lekner1991,
  title     = "Reflection and refraction by uniaxial crystals",
  author    = "Lekner, J",
  journal   = "J. Phys. Condens. Matter",
  publisher = "IOP Publishing",
  volume    =  3,
  number    =  32,
  pages     = "6121--6133",
  month     =  aug,
  year      =  1991
}

@inbook{Mondal2021,
  title     = "Fundamentals of natural fibres and textiles",
  editor    = "Mondal, Ibrahim H",
  publisher = "Woodhead Publishing",
  series    = "The Textile Institute Book Series",
  month     =  mar,
  year      =  2021,
  language  = "en",
  chapter = 2
}

@article{Huijts1994,
title = {The relation between molecular orientation and birefringence in PET and PEN fibres},
journal = {Polymer},
volume = {35},
number = {14},
pages = {3119-3121},
year = {1994},
issn = {0032-3861},
doi = {https://doi.org/10.1016/0032-3861(94)90429-4},
url = {https://www.sciencedirect.com/science/article/pii/0032386194904294},
author = {R.A. Huijts and S.M. Peters}
}

@inproceedings{Molnar2022,
author="Molnar, Christoph
and K{\"o}nig, Gunnar
and Herbinger, Julia
and Freiesleben, Timo
and Dandl, Susanne
and Scholbeck, Christian A.
and Casalicchio, Giuseppe
and Grosse-Wentrup, Moritz
and Bischl, Bernd",
editor="Holzinger, Andreas
and Goebel, Randy
and Fong, Ruth
and Moon, Taesup
and M{\"u}ller, Klaus-Robert
and Samek, Wojciech",
title="General Pitfalls of Model-Agnostic Interpretation Methods for Machine Learning Models",
bookTitle="xxAI - Beyond Explainable AI: International Workshop, Held in Conjunction with ICML 2020, July 18, 2020, Vienna, Austria, Revised and Extended Papers",
year="2022",
publisher="Springer International Publishing",
pages="39--68",
isbn="978-3-031-04083-2",
doi="10.1007/978-3-031-04083-2_4",
url="https://doi.org/10.1007/978-3-031-04083-2_4"
}

@inproceedings{
Frye2020,
title={Shapley explainability on the data manifold},
author={Christopher Frye and Damien de Mijolla and Tom Begley and Laurence Cowton and Megan Stanley and Ilya Feige},
booktitle={International Conference on Learning Representations},
year={2021},
url={https://openreview.net/forum?id=OPyWRrcjVQw}
}

@ARTICLE{Peng2005,
  author={Hanchuan Peng and Fuhui Long and Ding, C.},
  journal={IEEE Transactions on Pattern Analysis and Machine Intelligence}, 
  title={Feature selection based on mutual information criteria of max-dependency, max-relevance, and min-redundancy}, 
  year={2005},
  volume={27},
  number={8},
  pages={1226-1238},
  doi={10.1109/TPAMI.2005.159}}

@ARTICLE{Radovic2017,
  title     = "Minimum redundancy maximum relevance feature selection approach
               for temporal gene expression data",
  author    = "Radovic, Milos and Ghalwash, Mohamed and Filipovic, Nenad and
               Obradovic, Zoran",
  journal   = "BMC Bioinformatics",
  publisher = "Springer Science and Business Media LLC",
  volume    =  18,
  number    =  1,
  pages     = "9",
  month     =  jan,
  year      =  2017,
  copyright = "http://creativecommons.org/licenses/by/4.0/",
  language  = "en"
}

@misc{repository,
  title        = "{GitHub repository: Microfiber classification}",
  howpublished = "\url{https://github.com/JanAppelCZ/microfiber-classification}",
  year         = "2026",
  author        = "Jan Appel"
}

@ARTICLE{Zuo2013,
  title     = "Phase aberration compensation in digital holographic microscopy
               based on principal component analysis",
  author    = "Zuo, Chao and Chen, Qian and Qu, Weijuan and Asundi, Anand",
  journal   = "Opt. Lett.",
  publisher = "Optica Publishing Group",
  volume    =  38,
  number    =  10,
  pages     = "1724--1726",
  month     =  may,
  year      =  2013,
  copyright = "https://doi.org/10.1364/OA\_License\_v1\#VOR",
  language  = "en"
}

@ARTICLE{Behal2019,
  title     = "Quantitative phase imaging in common-path cross-referenced
               holographic microscopy using double-exposure method",
  author    = "B{\v e}hal, Jarom{\'\i}r",
  journal   = "Sci. Rep.",
  publisher = "Springer Science and Business Media LLC",
  volume    =  9,
  number    =  1,
  pages     = "9801",
  month     =  jul,
  year      =  2019,
  copyright = "https://creativecommons.org/licenses/by/4.0",
  language  = "en"
}

@ARTICLE{Sirico2022l,
  title     = "Compensation of aberrations in holographic microscopes: main
               strategies and applications",
  author    = "Sirico, Daniele Gaetano and Miccio, Lisa and Wang, Zhe and
               Memmolo, Pasquale and Xiao, Wen and Che, Leiping and Xin, Lu and
               Pan, Feng and Ferraro, Pietro",
  journal   = "Appl. Phys. B",
  publisher = "Springer Science and Business Media LLC",
  volume    =  128,
  number    =  4,
  month     =  apr,
  year      =  2022,
  copyright = "https://creativecommons.org/licenses/by/4.0",
  language  = "en"
}

@ARTICLE{Sanchez-Ortiga2014,
  title     = "Off-axis digital holographic microscopy: practical design
               parameters for operating at diffraction limit",
  author    = "S{\'a}nchez-Ortiga, Emilio and Doblas, Ana and Saavedra, Genaro
               and Mart{\'\i}nez-Corral, Manuel and Garcia-Sucerquia, Jorge",
  journal   = "Appl. Opt.",
  publisher = "Optica Publishing Group",
  volume    =  53,
  number    =  10,
  pages     = "2058--2066",
  month     =  apr,
  year      =  2014,
  copyright = "https://doi.org/10.1364/OA\_License\_v1\#VOR",
  language  = "en"
}

@ARTICLE{Hornik1989,
  title     = "Multilayer feedforward networks are universal approximators",
  author    = "Hornik, Kurt and Stinchcombe, Maxwell and White, Halbert",
  journal   = "Neural Netw.",
  publisher = "Elsevier BV",
  volume    =  2,
  number    =  5,
  pages     = "359--366",
  month     =  jan,
  year      =  1989,
  language  = "en"
}

@ARTICLE{Belkin2019,
  title     = "Reconciling modern machine-learning practice and the classical
               bias-variance trade-off",
  author    = "Belkin, Mikhail and Hsu, Daniel and Ma, Siyuan and Mandal,
               Soumik",
  journal   = "Proc. Natl. Acad. Sci. U. S. A.",
  publisher = "Proceedings of the National Academy of Sciences",
  volume    =  116,
  number    =  32,
  pages     = "15849--15854",
  month     =  aug,
  year      =  2019,
  copyright = "https://www.pnas.org/site/aboutpnas/licenses.xhtml",
  language  = "en"
}

@ARTICLE{Nakkiran2021,
  title     = "Deep double descent: where bigger models and more data hurt",
  author    = "Nakkiran, Preetum and Kaplun, Gal and Bansal, Yamini and Yang,
               Tristan and Barak, Boaz and Sutskever, Ilya",
  journal   = "J. Stat. Mech.",
  publisher = "IOP Publishing",
  volume    =  2021,
  number    =  12,
  pages     = "124003",
  month     =  dec,
  year      =  2021,
  copyright = "https://publishingsupport.iopscience.iop.org/iop-standard/v1"
}

@ARTICLE{Zhu2022,
  title     = "Microplastic pollution assessment with digital holography and
               zero-shot learning",
  author    = "Zhu, Yanmin and Lo, Hau Kwan Abby and Yeung, Chok Hang and Lam, Edmund Y",
  journal   = "APL Photonics",
  publisher = "AIP Publishing",
  volume    =  7,
  number    =  7,
  pages     = "076102",
  month     =  jul,
  year      =  2022,
  copyright = "https://creativecommons.org/licenses/by/4.0/",
  language  = "en"
}

@ARTICLE{Takahashi2023,
  title     = "Multimodal image and spectral feature learning for efficient
               analysis of water-suspended particles",
  author    = "Takahashi, Tomoko and Liu, Zonghua and Thevar, Thangavel and
               Burns, Nicholas and Lindsay, Dhugal and Watson, John and
               Mahajan, Sumeet and Yukioka, Satoru and Tanaka, Shuhei and
               Nagai, Yukiko and Thornton, Blair",
  journal   = "Opt. Express",
  publisher = "Optica Publishing Group",
  volume    =  31,
  number    =  5,
  pages     = "7492--7504",
  month     =  feb,
  year      =  2023,
  copyright = "https://creativecommons.org/licenses/by/4.0/",
  language  = "en"
}

@ARTICLE{Shi2022,
  title     = "Automatic quantification and classification of microplastics in
               scanning electron micrographs via deep learning",
  author    = "Shi, Bin and Patel, Medhavi and Yu, Dian and Yan, Jihui and Li,
               Zhengyu and Petriw, David and Pruyn, Thomas and Smyth, Kelsey
               and Passeport, Elodie and Miller, R J Dwayne and Howe, Jane Y",
  journal   = "Sci. Total Environ.",
  publisher = "Elsevier BV",
  volume    =  825,
  number    =  153903,
  pages     = "153903",
  month     =  jun,
  year      =  2022,
  copyright = "http://creativecommons.org/licenses/by-nc-nd/4.0/",
  language  = "en"
}

@ARTICLE{Guo2026,
  title     = "Computational polarimetric holography for efficient microplastic
               classification via a lightweight wavelet-enhanced vision
               transformer",
  author    = "Guo, Miao and Guo, Buyu and He, Shuangyan and Zhou, Zhou and
               Huang, Hui and Li, Peiliang",
  journal   = "Opt. Express",
  publisher = "Optica Publishing Group",
  volume    =  34,
  number    =  6,
  pages     = "9807--9827",
  month     =  mar,
  year      =  2026,
  language  = "en"
}

@ARTICLE{Huang2025,
  title     = "Computational polarized holography for automatic monitoring of
               microplastics in scattering aquatic environments",
  author    = "Huang, Jianqing and Zhu, Shuo and Li, Yuxing and Wang, Chutian
               and Lam, Edmund Y",
  journal   = "APL Photonics",
  publisher = "AIP Publishing",
  volume    =  10,
  number    =  3,
  month     =  mar,
  year      =  2025,
  copyright = "https://creativecommons.org/licenses/by/4.0/",
  language  = "en"
}

@ARTICLE{Zhu2024,
  title     = "Smart polarization and spectroscopic holography for real-time
               microplastics identification",
  author    = "Zhu, Yanmin and Li, Yuxing and Huang, Jianqing and Lam, Edmund Y",
  journal   = "Commun Eng",
  publisher = "Springer Science and Business Media LLC",
  volume    =  3,
  number    =  1,
  month     =  jan,
  year      =  2024,
  copyright = "https://creativecommons.org/licenses/by/4.0",
  language  = "en"
}

@ARTICLE{Saur2026,
  title         = "Classification of microplastic particles in water using
                   polarized light scattering and machine learning methods",
  author        = "Saur, Leonard and von Pawlowski, Marc and Gengenbach, Ulrich
                   and Sieber, Ingo and Shirali, Hossein and W{\"u}hrl, Lorenz
                   and Weng, Xiangyu and Kiko, Rainer and Pylatiuk, Christian",
  journal       =  "arXiv",
  month         =  mar,
  year          =  2026,
  copyright     =  "http://creativecommons.org/licenses/by-nc-sa/4.0/",
  archivePrefix =  "arXiv",
  primaryClass  = "cs.CV",
  eprint        = "2511.06901"
}

@ARTICLE{Yang2026,
  title     = "High-throughput microplastic differentiation using pixel-based
               polarization classification",
  author    = "Yang, Jianxiong and Li, Yuqing and Jiang, Feng and Liu, Hoi Man
               and Liu, Mengyang and Yan, Meng and Hu, Zheng and Han, Baohui
               and Zhang, Shoufeng and Chu, Xiaoting and Qiu, Zhigang and Liao,
               Ran and Ma, Hui",
  journal   = "ACS Photonics",
  publisher = "American Chemical Society (ACS)",
  volume    =  13,
  number    =  8,
  pages     = "2059--2071",
  month     =  apr,
  year      =  2026,
  language  = "en"
}

@ARTICLE{Montandon2024,
  title     = "Imaging‐based lensless polarization‐sensitive fluid stream
               analyzer for automated, label‐free, and cost‐effective
               microplastic classification",
  author    = "Montandon, Fraser and Nicolls, Fred",
  journal   = "Adv. Intell. Syst.",
  publisher = "Wiley",
  volume    =  6,
  number    =  12,
  month     =  dec,
  year      =  2024,
  copyright = "http://creativecommons.org/licenses/by/4.0/",
  language  = "en"
}

\end{document}